\documentclass[12pt,preprint]{aastex}

\usepackage{amsmath}
\usepackage{graphicx}
\usepackage{lscape}
\usepackage{color}

\shorttitle{Reprocessing of Ices in Protoplanetary Disks}
\shortauthors{Furuya et al.}

\begin{document}

\title{Reprocessing of Ices in Turbulent Protoplanetary Disks: Carbon and Nitrogen Chemistry}


\author{Kenji Furuya\altaffilmark{1,2} and Yuri Aikawa\altaffilmark{1}}
\email{furuya@strw.leidenuniv.nl}

\altaffiltext{1}{Department of Earth and Planetary Sciences, Kobe University, Kobe 657-8501, Japan}
\altaffiltext{2}{Leiden Observatory, Leiden University, P.O. Box 9513, 2300 RA, The Netherlands}



\begin{abstract}
We study the influence of the turbulent transport on ice chemistry in protoplanetary disks, focusing on carbon and nitrogen bearing molecules.
Chemical rate equations are solved with the diffusion term, mimicking the turbulent mixing in the vertical direction.
Turbulence can bring ice-coated dust grains from the midplane to the warm irradiated disk surface, 
and the ice mantles are reprocessed by photoreactions, thermal desorption, and surface reactions.
The upward transport decreases the abundance of methanol and ammonia ices at $r\lesssim30$ AU, 
because warm dust temperature prohibits their reformation on grain surfaces. 
This reprocessing could explain the smaller abundances of carbon and nitrogen bearing molecules in 
cometary coma than those in low-mass protostellar envelopes.
We also show the effect of mixing on the synthesis of complex organic molecules (COMs) are two ways:
(1) transport of ices from the midplane to the disk surface and (2) transport 
of atomic hydrogen from the surface to the midplane. The former enhances the 
COMs formation in the disk surface, while the latter suppresses it in the midplane.
Then, when mixing is strong, COMs are predominantly formed in the disk surface, 
while their parent molecules are (re)formed in the midplane.
This cycle expands the COMs distribution both vertically and radially outward compared with that in the non-turbulent model.
We derive the timescale of the sink mechanism by which CO and N$_2$ are converted to less volatile molecules to be depleted from the gas phase,
and find that the vertical mixing suppresses this mechanism in the inner disks.


\end{abstract}





\section{INTRODUCTION \label{coms:sec:intro}}
One of the most intriguing questions in the studies of astrochemistry is whether and how much of the pristine materials 
in the solar nebula were inherited from interstellar matter (ISM) \citep[see e.g., recent review by][]{caselli12,bergin13}.
Comets are thought to be the most pristine objects of the cold ice-bearing regions in the solar nebula \citep[e.g.,][]{brownlee03}.
When a comet approaches the sun, sublimation of volatile species produces the envelope around the cometary nuclei, which is called coma.
Many molecules have been detected in cometary coma, and their abundances with respect to water have been derived \citep[][and reference therein]{mumma11}.
All molecules detected in comets have also been detected in prestellar or protostellar cores.
This similarity suggests that cometary molecules could originate from the cloud/core phase.
\citet{oberg11} pointed out, however, that cometary abundances of CO, CH$_4$, and CH$_3$OH are lower 
than the median abundances of the ices in low-mass protostellar envelopes, obtained from the $Spitzer$ c2d Legacy ice survey.
\citet{ootsubo12} conducted a spectroscopic survey of cometary molecules using AKARI, and found that CO$_2$ is depleted in comets, as well.
In addition to carbon bearing molecules, NH$_3$ is also depleted in comets compared to the low-mass protostellar ices \citep[e.g.,][]{oberg11,kawakita11,biver12}.
Recent chemical-dynamical model suggests that the majority of water is delivered from cores to disks as ice \citep{visser11}.
If this was the case in the solar nebula, CH$_3$OH and NH$_3$ would be also delivered to the disk as ice,
since CH$_3$OH and NH$_3$ are co-desorbed with water as shown by temperature programmed desorption (TPD) experiments \citep{collings04}.
Then their depletions in comets may imply that some reprocessing of ices occurred in the solar nebula, 
although it is unclear whether the observed median composition well represents the ice composition in the protosolar envelope.

In this paper, we investigate the effect of turbulent mixing in the vertical direction on ice chemistry in Class II protoplanetary disks.
As demonstrated in our previous work \citep[][ here after Paper I]{furuya13}, the combination of upward transport of ices, thermal desorption, and 
photochemistry leads to destruction of ices.
This strongly contrasts with disk models without turbulence, where most of the ices survive for $>$10$^6$ yr, because they are mainly exist in the midplane, 
shielded from stellar UV and X-ray.
In Paper I, we focused on water ice and the HDO/H$_2$O ratio.
Here we focus on other simple icy species, such as NH$_3$ and CH$_3$OH.

In addition to simple molecules, we also investigate synthesis of complex organic molecules, such as HCOOCH$_3$, in turbulent disks.
One possible formation pathway of complex molecules is the grain surface reaction 
between heavy radicals formed via ultraviolet photolysis of simpler molecules \citep[e.g.,][and reference therein]{herbst09}.
Although formation of complex molecules in the gas phase has been also proposed \citep[e.g.,][]{charnley92}, 
recent experiments suggest that some key gas-phase reactions are less efficient than previously thought \citep{geppert06}.
For grain-surface formation, there are (at least) two important factors:
UV photons to form radicals and warm dust temperature \citep[e.g.,][]{oberg09c}.
In the cold temperature ($\sim$10 K), hydrogenation of atoms and molecules is the most efficient surface reaction.
When dust temperature is warm ($\gtrsim$30 K), surface chemistry involving heavy radicals becomes important, 
as they can be mobile on grain surfaces, while hydrogen atoms tend to evaporate rather than react \citep{garrod06,garrod08}.

Very recently, \citet{walsh14} studied spatial distributions of complex organic molecules in a disk without turbulent mixing.
Their model reasonably reproduces the observed abundance of the complex molecules in comets, 
by assuming relatively small diffusion energy barriers for icy species.
On the other hand, recent theoretical studies suggest that turbulent mixing could increase the formation efficiency of complex organic molecules \citep{semenov11,ciesla12}.
\citet{ciesla12} traced trajectories of dust grains in a turbulent disk, and showed that dust grains move throughout the disk, 
experiencing irradiation and warming.
They found that incident photon number on grain surfaces is sufficient to produce significant amount of complex organic molecules, 
if they assume the production efficiency of complex molecules per incident photon obtained from laboratory experiments.
It is thus important as a next step to calculate rate equations to evaluate the formation of complex organic molecules in turbulent disks.
While several authors have studied chemical evolution in turbulent disks numerically \citep[e.g.,][]{ilgner04,willacy06,hersant09,heinzeller11,
they have mainly focused on relatively simple gas phase molecules.}
To the best of our knowledge, \citet[][hereafter SW11]{semenov11} is the only work which investigated chemistry of complex organic molecules in turbulent disks by solving chemical rate equations, so far.
SW11 found that complex molecules, HCOOH ice and CH$_3$CHO ice, are much more abundant in the model with mixing than those in the model without mixing.

This paper is organized as follows.
In Section \ref{coms:sec:disk_structure}, we briefly describe physics and chemistry in our numerical model. 
In Section \ref{coms:sec:result}, we present our results, while in Section \ref{coms:sec:discussion}, 
we discuss the implication of our model results on the chemical compositions of comets.
We compare our results with SW11, and also discuss the uncertainties of ice photolysis and 
the effect of grain growth on our results.
We summarize our conclusions in Section \ref{coms:sec:conclusion}.


\section{PHYSICAL AND CHEMICAL MODELING} \label{coms:sec:disk_structure}
The physical and chemical models used in this work are almost the same as those in Paper I.
One dimensional reaction-diffusion equations \citep{xie95, willacy06} are solved in a disk physical model of \citet{nomura07}: 
\begin{equation}
\frac{\partial n_i}{\partial t} + \frac{\partial \phi_i}{\partial z} = P_i - L_i, \label{coms:eq:basic}
\end{equation}
\begin{equation}
\phi_i = -n_{\rm H} D_z\frac{{\partial }}{{\partial z}}\left(\frac{n_i}{n_{\rm H}}\right),
\end{equation}
where $n_i$, $P_i$, and $L_i$ represent the number density, production rate, and the loss rate, respectively, of species $i$.
The second term in the left-hand side of Equation (\ref{coms:eq:basic}) mimics turbulent mixing in the vertical direction, 
and $D_z$ is a diffusion coefficient.
Equation (\ref{coms:eq:basic}) is integrated for 10$^6$ yr using implicit finite differencing on a linear grid consisting of vertical 60 points at a specific radius.
In the rest of this section, we briefly outline our physical and chemical models.

\subsection{Disk structure} \label{coms:sec:phys_disk}
\begin{figure}
\epsscale{1.0}
\plotone{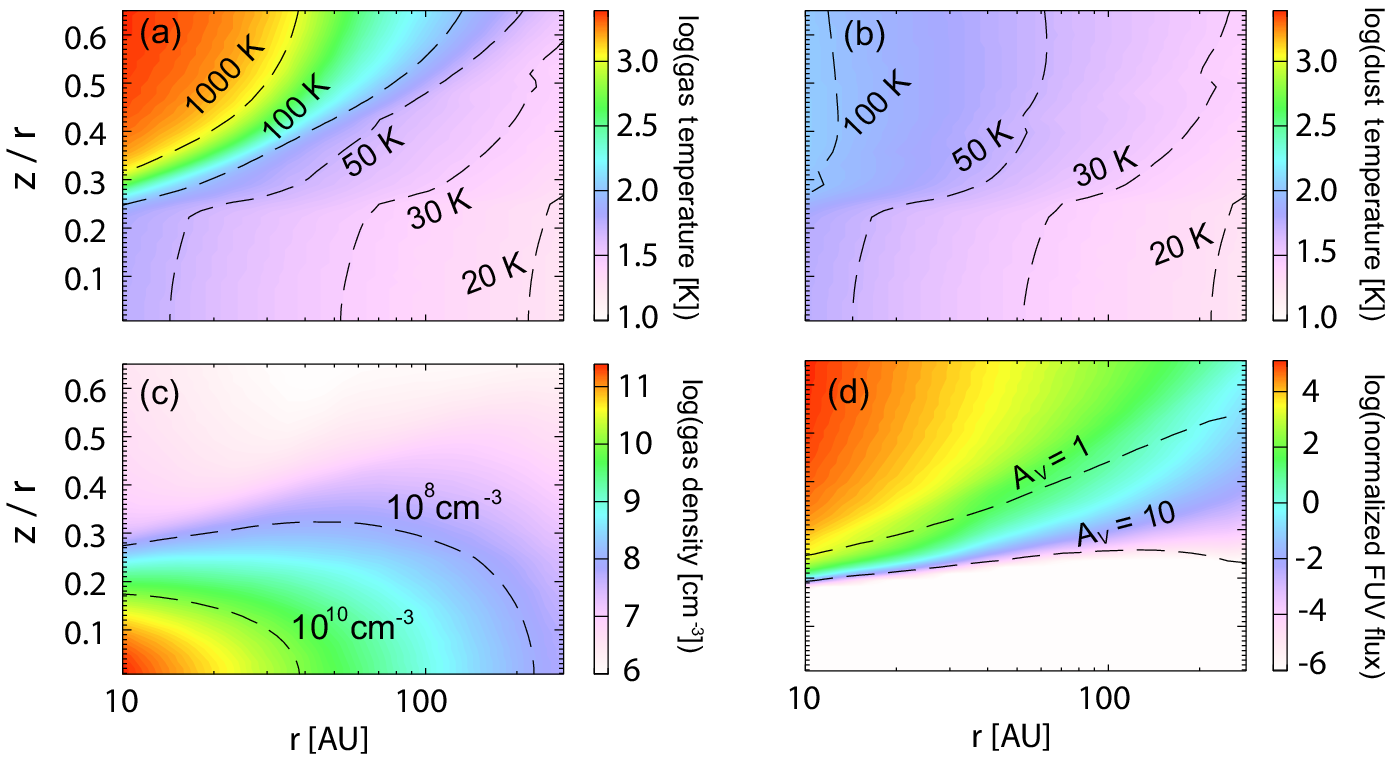}
\caption{Spatial distributions of the gas temperature (a), dust temperature (b), number density of gases (c), 
and wavelength-integrated FUV flux normalized by Draine field \citep[$1.6\times10^{-3}$ erg cm$^{-2}$ s$^{-1}$;][(d)]{draine78}.
In panel (d), the dashed lines indicate the height at which the vertical visual extinction of interstellar radiation field reaches unity and ten.
The vertical axes represent height normalized by the radius.
\label{coms:fig:phys}}
\end{figure}

We adopt a disk model of \citet{nomura07}; it is a steady, axisymmetric Keplerian disk surrounding a typical T Tauri star 
with the mass $M_{*}=0.5$ $M_{\bigodot}$, radius $R_{*}=2$ $R_{\bigodot}$, and effective temperature $T_{*}=4000$ K.
The stellar UV and X-ray luminosities are set to be $10^{31}$ erg s$^{-1}$ and $10^{30}$ erg s$^{-1}$, respectively, 
based on the spectrum observed towards TW Hya \citep[e.g.,][]{herczeg02,kastner02}.
We note that X-ray ionization dominates over cosmic ray ionization in the disk surface (see Figure 7 in \citet{nomura07}).
The X-ray luminosity adopted in our model is higher than the median value in T Tauri stars \citep[$\sim 10^{29.5}$ erg s$^{-1}$;][]{preibisch05}, 
while the X-ray spectrum is soft \citep{kastner02}.
X-ray ionization rate is proportional to X-ray flux, while X-ray penetrates deeper into the disk if the spectrum is harder \citep{verner95}.
The dust-to-gas mass ratio is set to be 0.01 with the dust size distribution model adequate for dense clouds \citep{weingartner01}. 
The gas temperature, dust temperature and density distributions of the disk are calculated self-consistently, 
considering various heating and cooling mechanisms \citep[see][for details]{nomura05,nomura07}.
The disk structure adopted in this work is shown in Figure \ref{coms:fig:phys}.

We assume that the origin of the disk turbulence is the magnetorotational instability \citep{balbus91}.
The vertical diffusion coefficient is assumed to be
\begin{equation}
D_z = \alpha_z c_s^2/\Omega,
\end{equation}
where $\Omega$ and $c_s$ are the Keplerian orbital frequency and local sound speed, respectively \citep[][]{fromang06, okuzumi11}.
We can safely use the same $D_z$ for gaseous and ice mantle species, 
since dust radius is assumed to be 0.1 $\mu$m in our chemical model and the ratio of gas to dust diffusivity of such a small grain is nearly unity.
We assume that the alpha parameter, $\alpha_z$, is constant for simplicity, and run three models with $\alpha_z = 0$, $10^{-3}$, and $10^{-2}$.

\subsection{Chemical Model}
To compute disk chemistry, a two-phase model, which consists of gas-phase and grain-surface species, 
is adopted \citep{hasegawa92}.
Our chemical reaction network is based on \citet{garrod06}, and modified to be applicable to disk chemistry.
The species with chlorine and phosphorus, molecules with more than four carbon atoms, and their relevant reactions are excluded from the network
to reduce the computational time.
As chemical processes, we take into account the gas phase reactions, interaction between gas and grains, and grain surface reactions.
Since the detailed explanation of our chemical model can be found in Paper I, here we only describe updates of our reaction network, 
photochemistry of ices, and the initial abundance.

The reaction network and adopted parameters are the same as in a fiducial model of Paper I, except for the following three updates.
Firstly, desorption energy of atomic hydrogen is set to be 600 K instead of 450 K; 
the former is adequate for amorphous water ice, while the latter is adequate for crystalline water ice \citep{al-halabi07}.
Secondly, the activation energy barrier of CO + OH reaction on grain surfaces are set to be 176 K \citep{cuppen09} instead of 80 K \citep{ruffle01}.
The value of this barrier is currently controversial;
\citet{oba10} concluded that the reaction proceeds with little or no barrier from their laboratory experiments,
while \citet{noble11} concluded that the barrier is higher than previously thought.
In this work, we adopt the barrier of the same reaction (CO + OH) in the gas phase.
Thirdly, photodissociation branching ratios of methanol ice are updated following the experimental work of \citet{oberg09c}.
CH$_3$OH is a parent molecule of more complex species, and the photodissociation branching ratio is an important parameter 
for the formation efficiency of complex organic molecules \citep{laas11}.
\citet{oberg09c} estimated CH$_3$OH photodissociation branching ratio (CH$_2$OH+H):(CH$_3$O+H):(CH$_3$+OH) of 5:1:$<$1, based on their experiments.
Since we do not discriminate between isomers, CH$_2$OH and CH$_3$O, in our network, the branching ratio (CH$_3$O+H):(CH$_3$+OH) is set to be 9:1.

UV irradiation experiments have shown that ice molecules can be photodissociated and photodesorbed \citep[e.g.,][]{gerakines96,westley95}.
Photodissociation of ice molecules produces radicals, while photodesorption can keep a fraction of molecules in the gas phase even at low dust temperatures.
The photodissociation rates (cm$^{-3}$ s$^{-1}$) of ice species are calculated as follows:
\begin{align}
R_{{\rm phs}, i} & = \pi a^2n_{\rm d} \int_{}^{} 4\pi J(\lambda)[1-\exp(-\tau_i(\lambda))] d\lambda, \label{coms:eq:rphs}\\
\tau_i(\lambda) &= \frac{N_{\rm p}\sigma_i(\lambda)n_i}{\sigma_{\rm site}n_{\rm ice}},\label{coms:eq:tau_i}\\
N_{\rm p} & = {\rm min}(N_{\rm layer}, 2), \label{coms:eq:np}
\end{align}
where $a$, $n_{\rm d}$, $n_{\rm ice}$, $\sigma_{\rm site}$, and $N_{\rm layer}$ are the grain radius, 
the number density of the grain, the total number density of ice mantle species per unit volume of gases, the site area, and
the number of monolayer of ice mantles ($n_{\rm ice}/n_{\rm d}N_{\rm site}$, with the number of adsorption sites on a grain $N_{\rm site} \sim 10^6$), respectively.
While $\tau_i$ represents the optical depth of ice mantle species $i$, the term $(1-\exp(-\tau_i))$ represents the fraction of photon absorbed by species $i$ 
in the ice mantle.
$J(\lambda)$ is the mean intensity of the radiation field measured in units of photons cm$^{-2}$ s$^{-1}$ \AA $^{-1}$ str$^{-1}$ at the wavelength of $\lambda$. 
We consider stellar, interstellar, and X-ray-induced and cosmic-ray-induced UV radiation.
The cross sections, $\sigma_i(\lambda)$, for water and CO$_2$ ices are taken from \citet{mason06}.
For the other species, we use the same cross section used for corresponding gaseous species,
because data are not available in the literature.
The UV radiation can penetrate into deep layers of ice mantle.
But we assume that only the uppermost monolayers can be dissociated as the outcome of photoabsorption, following Paper I.
The photoproducts in deeper layers are assumed to recombine immediately.
The number of active layer for photodissociation, $N_{\rm p}$, is restricted to less than two for all species, considering surface roughness.
We discuss the effect of these assumptions on our results in Section \ref{coms:dis:phd}.
Equations (\ref{coms:eq:rphs}--\ref{coms:eq:np}) are essentially the same as Equations (9--11) in Paper I when $\tau_i$ is much less than unity.
This is the case for all species in our fiducial models, since $(\sigma_i/\sigma_{\rm site})$ and/or $(n_i/n_{\rm ice})$ are much less than unity. 

The photodesorption rates are calculated as \citep[e.g.,][]{visser11}
\begin{align}
R_{{\rm pd}, i}(r, z) &= \pi a^2 n_{\rm d}\theta_i F_{\rm FUV}(r, z)Y_i, \label{coms:eq:pd} \\
\theta_i &= \frac{n_i}{{\rm max}(n_{\rm ice}, n_{\rm d}N_{\rm site})}, \label{coms:eq:theta}
\end{align}
where $F_{\rm FUV}$ are the FUV photon number flux integrated in the range of 912--2000 {\AA}.
We use photodesorption yields per incident FUV photon, $Y_i$, derived from experimental work for H$_2$O, CO$_2$, CO, O$_2$, and N$_2$ ices \citep[][]{oberg09a,oberg09b,fayolle11,fayolle13}.
While the yields vary among species, and depend on the shape of UV spectrum, a typical value is 10$^{-3}$--10$^{-2}$.
We set the yield for the species without laboratory data to be 10$^{-3}$ for simplicity.
We also take into account non-thermal desorption via stochastic heating by high-energy particles \citep{hasegawa93} and reactive desorption \citep{garrod07}.

Initial molecular abundance in the disk (see Table \ref{coms:table:initial} for selected species) is the same as in Paper I, and was obtained by 
calculating the molecular evolution in a star forming core model of \citet{masunaga00} \citep[see also][]{aikawa08}.
We adopt the so-called low metal values as the elemental abundances \citep[see Table 1 of][]{aikawa01}. 
We integrate the rate equations (similar to Equation (\ref{coms:eq:basic}) without the diffusion term, but with Lagrange derivative) 
along the trajectory of an infalling fluid parcel, 
which was at the radius of 10$^4$ AU in the initial prestellar core and reaches 60 AU at the protostar age of $9.3\times10^4$ yr,
which is the final time of the core simulation \citep{masunaga00,aikawa08}.
The abundances in this protostellar core model are mostly constant at $r < 100$ AU, where molecules are mainly in the gas phase.


\section{RESULT\label{coms:sec:result}}
Here, we describe ice chemistry in turbulent protoplanetary disks.
We focus on simple species in Section \ref{coms:sec:simple}, while results of complex organic molecules are presented in Section \ref{coms:sec:complex}.
Upward transport of ices by turbulence also affects the spatial distributions of gas-phase molecules through thermal and non-thermal desorption, 
which will be discussed in Section \ref{coms:sec:comparison_obs}.
CO ans N$_2$ can be converted to less volatile species and depleted from the gas phase.
The mechanism and timescale of such sink effect are analyzed in Section \ref{coms:sec:depletion}.
In the following we discuss the results at 10$^6$ yr, which is the typical age of T Tauri stars with circumstellar disks, unless otherwise stated.
In Table \ref{coms:table:tevap}, we list evaporation temperature of assorted molecules when the gas density is 10$^6$--10$^{12}$ cm$^{-3}$.
A combination of Table \ref{coms:table:tevap} and Figure \ref{coms:fig:phys}(b) tells us in which disk regions the thermal desorption 
is important for each molecular species (see also Appendix A). 

\subsection{Simple Saturated Molecules: CH$_3$OH and NH$_3$\label{coms:sec:simple}}
The left panels in Figure \ref{coms:fig:ab_simple} show the resultant fractional abundances of selected ice-mantle species 
with respect to hydrogen nucleus in the model without mixing.
Although molecules are mostly in the gas phase in our initial abundances,
they are adsorbed onto grains in a short timescale ($\sim$10$^9/n$ yr) in regions colder than their sublimation temperatures.
In the following, we compare the ice abundances at 10$^6$ yr to the ice abundances right after the initial freeze out,
which are essentially the same as the initial gas-phase abundance.

H$_2$CO ice abundance is low compared to the initial value ($1.5\times10^{-5}$) throughout the disk; 
H$_2$CO ice is converted to CH$_3$OH ice via subsequent hydrogenation. 
In the regions where dust temperature ($T_{\rm d}$) is higher than 40 K, H$_2$CO is desorbed to the gas phase and destroyed by ion-neutral reactions.
CH$_3$OH and NH$_3$ ices show similar spatial distributions; 
they are abundant ($\sim10^{-6}$--10$^{-5}$) near the midplane, 
while the abundance is low ($\lesssim$10$^{-8}$) in the upper layers because of the strong stellar UV radiation.
Solid lines in Figure \ref{coms:fig:r_vs_sigma_simple} shows the molecular column densities normalized by that of water ice in the model without mixing.
Note that water-ice column density stays constant for 10$^6$ yr in this model, i.e., the water ice abundance near the midplane is $\sim$10$^{-4}$ relative to hydrogen nuclei.
The column density ratios of CH$_3$OH and NH$_3$ ices to water ice are higher than their initial values regardless of a distance from a central star.
This is due to their efficient formation in the disk.
Once formed, these ices are not destroyed efficiently in the regions where UV photons and X-rays are attenuated.
The column density ratio of CH$_3$OH ice to water ice reaches the maximum at $r\gtrsim 120$ AU, where dust temperatures are lower than 
the sublimation temperature of CO.

Turbulence brings ices to the warm irradiated disk surface, where ices are destroyed via photodissociation and/or photodesorption by the stellar UV photons.
On the other hand, gaseous species, such as atomic oxygen, nitrogen, and CO, in the disk surface are transported to the midplane to 
(re)form the ices and/or other gaseous species. 
This cycle changes the ice distributions significantly.
The abundances of CH$_3$OH ice and NH$_3$ ice decrease with time at $r\lesssim30$ AU,
since hydrogenation on grain surfaces (i.e., reformation of the saturated ices) is prohibited by high dust temperatures even near the midplane.
Instead, abundances of CO and N$_2$ in the gas phase increase with time.
In contrast to CH$_3$OH ice and NH$_3$ ice, water ice abundance decreases only at $r\lesssim 20$ AU.
At $r\sim$20--30 AU the resultant ice composition is methanol-poor and ammonia-poor compared to the initial composition.
The critical temperature (and thus radius) for reformation varies among molecules due to their formation paths.
Methanol ice is formed via the reactions with activation barriers, while water ice is mainly formed via barrierless reactions.
On the other hand, NH$_3$ ice formation is restricted by the destruction of N$_2$ in the gas phase to form atomic N.

As shown in Paper I, the destruction timescale of water ice is limited by their transport timescale in the vertical direction, 
i.e., $\tau^{\rm des}_{\rm ice} = N_{\rm ice}/\phi^{\rm crit}_{\rm ice}$, 
where $N_{\rm ice}$ is the ice column density, and $\phi^{\rm crit}_{\rm ice}$ is the flux at a layer above which destruction of a molecule via photoreactions is sufficiently fast.
We found destruction timescale of CH$_3$OH ice and NH$_3$ ice are similar to that of water ice, e.g., $\tau^{\rm des}_{\rm ice}\sim6(40)\times10^4$ yr at $r=10$ AU, 
if $\alpha_z=10^{-2}(10^{-3})$ (Equation (33) in Paper I).
Note that $\tau^{\rm des}_{\rm ice}$ is comparable to the accretion timescale, $\tau_{\rm acc} = r/v_{\rm acc}$ (Paper I).
At $r>30$ AU, lower dust temperatures allow efficient reformation of CH$_3$OH and NH$_3$ ices.
Turbulence continuously brings them to the disk surface, where destruction and supply from the midplane of the ices are balanced; 
the ices are abundant even at $A_V=1$ mag from the disk surface, if $\alpha_z=10^{-2}$.


\subsection{Complex Organic Molecules \label{coms:sec:complex}}
When dust temperature is warm ($\gtrsim$30 K), radical-radical association reactions become important, 
as radicals can be mobile on grain surfaces, while hydrogen atoms tend to evaporate rather than react with radicals \citep{garrod06,garrod08}.
We focus on HCOOCH$_3$ and CH$_3$OCH$_3$ as representatives of complex organic molecules, 
since they are often detected in high-mass and low--mass protostellar envelopes \citep[e.g.,][]{nummelin00,cazaux03}.
Regardless of the mixing strength, HCOOCH$_3$ ice and CH$_3$OCH$_3$ ice mainly form via the following grain-surface reactions in our models: 
\begin{align}
{\rm HCO} + {\rm CH_3O} & \rightarrow {\rm HCOOCH_{3}}, \label{coms:react:hcooch3}\\
{\rm CH_{3}} + {\rm CH_3O} & \rightarrow {\rm CH_3OCH_{3}}. \label{coms:react:ch3och3}
\end{align}

The left panels of Figure \ref{coms:fig:ab_com} shows the spatial distributions of abundances of HCOOCH$_3$ ice and CH$_3$OCH$_3$ ice in the model without mixing.
These complex species are moderately abundant (10$^{-8}$--10$^{-7}$) near the midplane in specific radii; 
HCOOCH$_3$ ice is most abundant in the midplane of $r=20$--30 AU, while CH$_3$OCH$_3$ ice is abundant in the outer regions ($r=30$--90 AU).
At $r\lesssim20$ AU, even heavy radicals tend to evaporate rather than react, while at $r\gtrsim90$ AU, hydrogenation of the radicals to form simple molecules becomes more efficient.

Figure \ref{coms:fig:t_vs_ab_nd} shows the temporal variations of their abundances in the midplane at $r=25$ AU and 80 AU in the model without mixing.
At $r = 25$ AU, HCOOCH$_3$ ice abundance increases in the timescale of several 10$^4$ yr.
CH$_3$O radical is mainly formed via the hydrogenation of H$_2$CO on grain surfaces, 
while HCO radical is mainly formed by the surface reaction of 
\begin{equation}
{\rm H_2CO} + {\rm OH} \rightarrow {\rm HCO} + {\rm H_2O}. \label{coms:eq:h2co+oh}
\end{equation}
At this radius, OH radical is mainly formed by the following pathway: 
${\rm CO}  \xrightarrow[]{{\rm He^+}} {\rm O}  \xrightarrow[]{{\rm grain}} {\rm O_{ice}} \xrightarrow[]{{\rm HNO_{ice}}} {\rm OH_{ice}}$.
We note it produces OH more efficiently than photodissociation of water ice by cosmic-ray induced photons, which is the dominant UV source in the midplane,
in our fiducial models.
If we assume $N_{\rm p}=N_{\rm layer}$ in Equation (\ref{coms:eq:tau_i}), OH production via water photodissociation becomes as efficient as O + HNO.
At $r\gtrsim 30$ AU, hydrogenation of HCO ice suppress the HCOOCH$_3$ ice formation.
On the other hand, at $r = 80$ AU, CH$_3$OCH$_3$ ice abundance increases at $t=2\times10^5$ yr,
when HNO ice abundance falls.
Until then, HNO is the main reaction partner of CH$_3$ \citep{garrod06}.
At $r\lesssim30$ AU, the fall of HNO ice abundance does not occur in 10$^6$ yr, 
probably because the higher dust temperatures suppress the conversion of HNO to NH$_3$.
These explain the difference in the spatial distributions of HCOOCH$_3$ and CH$_3$OCH$_3$ ices; 
HNO ice enhances the formation of the former, but suppresses the formation of the latter.

Effect of turbulent mixing on the formation of complex organic molecules are twofold:
(1) transport of ices from the midplane to the disk surface and
(2) transport of atomic hydrogen from the disk surface to the midplane.
The former enhances the formation of complex molecules in the disk surface, 
while the latter enhances the hydrogenation of radicals in the midplane to suppress the formation of complex molecules (but see Section \ref{coms:sec:comet}).
As a result, complex organic molecules are mainly formed in the region with $A_{\rm V} =$ a few mag, and turbulence brings them to the deeper layers into the disk. 
It implies that in turbulent disks, COMs can be synthesized even in the outer disk where dust temperatures in the midplane are 
too low for radicals to diffuse on grain surfaces, since dust temperatures are higher in the disk surface.
In the model with $\alpha_z=10^{-2}$, the abundances of HCOOCH$_3$ ice and CH$_3$OCH$_3$ ice reach the maximum at $r=60$ AU.
In the inner disk surface, radicals tend to desorb thermally rather than react, while in the outer disk surface, hydrogenation of radicals becomes more efficient.

Figure \ref{coms:fig:t_vs_ab_sd} shows temporal variations of the abundances of the complex species in the model with $\alpha_z = 10^{-2}$ 
at various disk height at $r=60$ AU.
In the disk surface, CH$_3$, CH$_3$O, and OH radicals are mainly formed by the photodissociation of CH$_3$OH and water on grain surfaces, 
while HCO ice is mainly formed by Reaction (\ref{coms:eq:h2co+oh}).
In the disk surface, HCOOCH$_3$ ice is temporally abundant ($\sim10^{-7}$) at $t \lesssim 5\times10^3$ yr, when H$_2$CO ice is still abundant.
After $t\sim10^5$ yr, the abundances of the complex species do not significantly vary with time in the disk surface.
The nearly steady-state values of $\sim$3$\times 10^{-8}$ are determined by the the balance between the destruction rate by photoreactions and the formation rate. 
The latter is determined by the competition between radical-radical reactions and hydrogenation of radicals, 
the rates of which are dependent on dust temperature and the abundance of atomic hydrogen.
The terminal abundances of complex molecules are not high compared to the maximum abundances in the model without mixing 
(e.g., $8\times10^{-7}$ for HCOOCH$_3$ and $5\times10^{-8}$ for CH$_3$OCH$_3$), 
despite much larger radical formation rates in the disk surface, because the abundance of atomic hydrogen is also high in the disk surface.
On the other hand, in the midplane, the abundances of the complex molecules increase with time in the timescale of $\sim$10$^5$ yr,
which is comparable to the transport timescale of the forming molecules from the disk surface to the midplane, $L^2/D_z$, 
where $L$ is the distance between the disk surface ($A_V\sim1$ mag) and midplane.
After $t\gtrsim10^6$ yr, the abundances in the midplane and disk surface are almost the same.
In the case of $\alpha_z=10^{-3}$, in which the transport timescale is longer ($\sim$10$^6$ yr at $r=60$ AU), 
disk surface is slightly abundant in complex molecules compared to the midplane at $t=10^6$ yr.

Figure \ref{coms:fig:r_vs_sigma_com} shows the column densities of the complex species normalized by that of CH$_3$OH ice,
which is a parent molecule of the complex species.
In the model without mixing, the maximum column density ratios for HCOOCH$_3$ ice and CH$_3$OCH$_3$ ice are 0.05 and 10$^{-3}$, respectively,
which are much higher than the initial values ($4\times10^{-4}$ and $10^{-5}$).
In the model with mixing, the maximum column density ratios are smaller than in the model without mixing,
again because of abundant atomic hydrogen both in the disk surface and in the midplane.

Our results suggest that the main formation site of complex organic molecules in disks depends on the turbulent strength.
In the disk with weak turbulence ($\alpha_z \leq 10^{-3}$), they are mainly formed near the midplane. 
On the other hand, when mixing is strong ($\alpha_z =10^{-2}$), they are mainly formed in the warm irradiated disk surface, 
while their parent molecules, such as methanol, are (re)formed in the cold midplane.
This cycle expands the disk regions with moderately abundant COMs abundance ($x_i \gtrsim 10^{-8}$) 
both in the vertical direction and outer disk radii compared with that in the model without mixing. 
We will discuss the effect of grain growth on these results in Section \ref{coms:dis:grain}.)

The two cases may be distinguished by deuterium-to-hydrogen (D/H) ratios of the complex molecules.
Although we do not show their D/H ratios in the current work, in the former case the D/H ratios should be similar to the molecular D/H ratio in the initial disk;
if the initial disk inherits the high molecular D/H ratio of cold cloud cores, the complex molecules formed there would also be highly deuterated.
In the case of strong turbulence, on the other hand, their D/H ratios would be lower than the interstellar values, since temperature in the disk surface 
is higher than that in the cloud core (Paper I).


\subsection{Gas-phase species \label{coms:sec:comparison_obs}}
Observational study of ices in disks is possible at least for water \citep{terada07,honda09},
but is often not straightforward because of the contamination of foreground components and geometry of the objects \citep[e.g.,][]{pontoppidan05,aikawa12}.
Ice mantle species are desorbed to the gas phase thermally and/or non-thermally.
It is possible to access the ice chemistry in disks thorough observing emission lines of the gas-phase species \citep{hogerheijde11}.
In this subsection we analyze the gas phase abundance of H$_2$CO and CH$_3$OH.
H$_2$CO is one of the most complex molecules detected in protoplanetary disks.
CH$_3$OH has not yet been detected in the disks, but is obviously an important target in ongoing ALMA observations.

The transport of ices affects the abundances of gaseous species in the disk surface through photodesorption, 
while transport of the gas from the upper layers keeps a fraction of CH$_3$OH and H$_2$CO in the gas phase ($n_i/n_{\rm H}=x_i<10^{-12}$) in 
the cold midplane \citep[see e.g.,][]{semenov06,aikawa07}.
Figure \ref{coms:fig:r_vs_sigma_gas} shows that in the model with $\alpha_z=10^{-2}$, gaseous CH$_3$OH is abundant in the disk surface ($A_V \sim 2$ mag) 
of $r\lesssim200$ AU.
There, CH$_3$OH abundance is determined by a local balance between photodissociaition and photodesorption:
\begin{equation}
F_{\rm FUV}\sigma_{\rm Ly\alpha} n_{\rm CH_3OH}^{\rm g} = F_{\rm FUV} \pi a^2 \left(\frac{n_{\rm CH_3OH}^{\rm s}}{n_{\rm ice}}\right) Y n_{\rm d},\label{coms:eq:balance}
\end{equation}
if dust grains are covered by more than one monolayer of ice (i.e., $n_{\rm ice} \ge n_{\rm d}N_{\rm site} \sim 10^{-6}n_{\rm H}$, 
see Equations (\ref{coms:eq:pd}) and (\ref{coms:eq:theta})).
The parameter $\sigma_{\rm Ly\alpha}$ is the photodissociation cross section of methanol 
at Ly$\alpha$ wavelength \citep[$1.4\times 10^{-17}$ cm$^2$;][]{vandishoeck06}, while 
$n_{\rm CH_3OH}^{\rm g}$ and $n_{\rm CH_3OH}^{\rm s}$ are the number density of methanol in the gas phase and in ice mantles, respectively.
We found that the photodesorption rate is about twice higher than the rate of reactive desorption 
if $\sim$1 \% of the (re)produced methanol ice is desorbed.
From Equation (\ref{coms:eq:balance}), the gaseous CH$_3$OH abundance is 
\begin{equation}
x^{\rm g}_{\rm CH_3OH} = 6\times 10^{-9}\left(\frac{n_{\rm CH_3OH}^{\rm s}/n_{\rm ice}}{0.1}\right)\left(\frac{\pi a^2 x_d}{6\times10^{-22}\,{\rm cm^2}}\right)
\left(\frac{Y}{10^{-3}}\right)\left(\frac{\sigma_{\rm Ly\alpha}}{10^{-17}\,{\rm cm^2}}\right)^{-1}.\label{coms:eq:x_sd}
\end{equation}
It is clear that the abundance is independent of UV photon flux and number density of the gas, 
but is dependent on ice compositions.
A similar analysis was previously performed for water vapor in disks \citep{dominik05} and in molecular clouds \citep{hollenbach09}.
At $r \gtrsim 200$ AU, on the other hand, ion-neutral reactions are the dominant destruction path of gaseous CH$_3$OH; 
the abundance is weakly dependent on number density of the gas ($x^{\rm g}_{\rm CH_3OH} \propto n^{-0.5}$).
The threshold radius should depend on the stellar UV and X-ray luminosity.

While CH$_3$OH is efficiently formed only on grain surfaces, H$_2$CO is mainly formed by the gas-phase reaction of CH$_3$ + O \citep[e.g.,][]{dalgarno73,aikawa99,walsh14}.
Nevertheless gas-phase H$_2$CO abundance is also enhanced in the disk surface (Figure \ref{coms:fig:r_vs_sigma_gas}), 
because CH$_3$ ice is formed by the photodissociation of CH$_3$OH ice and evaporates at $T_{\rm d}\gtrsim25$ K.

Radial profiles of the column densities of gaseous H$_2$CO and CH$_3$OH are shown in the right panels of Figure \ref{coms:fig:r_vs_sigma_gas}.
In the model without mixing, they increase with radius except for H$_2$CO at $r \lesssim 30$ AU.
When mixing is considered, the column densities are greatly enhanced at $r\gtrsim30$ AU, while they decrease in the inner disks (see Section \ref{coms:sec:simple}).
Then the radial column density profiles are ring-like in the models with $\alpha_z=10^{-2}$ as recently suggested by the SMA observations of H$_2$CO in 
Herbig Ae star HD 163296 \citep{qi13}.
In our model, the inner edge of the ring is determined by the efficiency of CO hydrogenation on grain surfaces.
We found analytical expressions of the column density of gaseous CH$_3$OH, which is depicted by dot-dashed lines in Figure \ref{coms:fig:r_vs_sigma_gas}.
The analytical expressions and their derivation are given in Appendix B.

As in the case of methanol, photodesorption is the main mechanism to desorp complex organic species into the gas phase in our models.
The maximum abundances ($10^{-12}$--10$^{-11}$) and column densities ($\sim$10$^{10}$ cm$^{-2}$) are not high even in the model with $\alpha_z=10^{-2}$, 
since they are not dominant species in ices.
Our model predicts that more complex molecules than CH$_3$OH in the gas phase, such as CH$_3$OCH$_3$, are much less abundant than CH$_3$OH and H$_2$CO.
This result agrees with \citet{walsh14}.

\subsection{Carbon and nitrogen depletion in the gas phase \label{coms:sec:depletion}}
CO freeze-out in the midplane with $T_{\rm d} \lesssim 20$--25 K is now established in both theoretical models and observations \citep{aikawa96,qi11,qi13}.
In theoretical models, however, CO could be depleted from the gas phase at higher dust temperatures;
carbon is locked in ices as carbon-chain molecules and CO$_2$ in the regions where $T_{\rm d}$ is 
higher than the sublimation temperature of CO, but lower than that of the latter species.
It is called carbon sink \citep{aikawa97}.
Recently, \citet{favre13} found that CO abundance in the warm layer ($T>20$ K) of TW hya is significantly lower ($(0.1-3)\times10^{-5}$) 
than the canonical value of $10^{-4}$;
this may imply that sink mechanism is really working in TW hya.
The model of \citet{aikawa97} showed that nitrogen is also locked in ices as NH$_3$ in the regions where $T_{\rm d} > 25$ K.

Figure \ref{coms:fig:sink_2d} shows the spatial distributions of abundances of CO and N$_2$ in our models.
In the model without mixing, we can see that CO and N$_2$ in the gas phase are significantly depleted even at $t=10^5$ yr in a layer at $z/r \sim 0.2$--0.4,
where stellar UV is heavily attenuated, but ionization rate is high due to X-ray.
At $t=10^6$ yr, carbon and nitrogen are significantly depleted throughout the disk, except for the irradiated disk surface and warm inner midplane.
To understand carbon and nitrogen chemistry more deeply, we analyze the timescale of the sink mechanisms and the effect of turbulent mixing in this subsection.

\subsubsection{Sink Mechanisms}
There are two possible mechanisms to convert CO into less volatile species: gas-phase route and grain-surface route.
Timescale of the gas-phase route is limited by the destruction timescale of CO by helium ion: 
\begin{equation}
{\rm CO} + {\rm He^+} \rightarrow {\rm C^+} + {\rm O}. \label{coms:react:co+he+}
\end{equation}
Carbon ion reacts with other carbon-bearing species, such as CH$_4$, followed by recombination with an electron or a negatively charged grain to form carbon-chain molecules.
On the other hand, atomic oxygen is adsorbed onto dust grains and is converted to CO$_2$ and other O-bearing species (Figure \ref{coms:fig:r_vs_sigma_simple}).
Helium ion is produced by the ionization of atomic helium, and destroyed mainly by CO (Reaction (\ref{coms:react:co+he+}))
as long as $x_{\rm CO} \gtrsim 0.5x_{\rm N_2}$ and $x_{\rm CO} \gtrsim 5\times10^{-6}x_{\rm H_2}$.
Then characteristic timescale of carbon depletion, $\tau_{\rm CO}$, is 
\begin{equation}
\tau_{\rm CO} = \frac{1}{\beta x_{\rm He^+}} \approx \frac{x_{\rm CO}}{\xi_{\rm He} x_{\rm He}} \approx 5\times 10^{5}\left(\frac{x_{\rm CO}}{3.5\times 10^{-5}}\right)
\left(\frac{\xi_{\rm He}}{2.5\times 10^{-17}\,{\rm s^{-1}}}\right)^{-1} \,\,{\rm yr}, \label{coms:eq:t_co}
\end{equation}
where ${\xi_{\rm He}}$ and $\beta$ are the ionization rate of He atom and the rate coefficient of Reaction (\ref{coms:react:co+he+}), respectively.
It should be noted that $\tau_{\rm CO}$ is proportional to CO abundance.
In other words, once CO abundance drops after $t>\tau_{\rm CO}$, sink effect is accelerated.
We can see this behavior in Figure \ref{coms:fig:sink}, which shows temporal variations of fractions of elemental carbon and nitrogen 
in the form of selected species at the midplane of $r=25$ AU, where $T_{\rm d} \sim 40$ K, in the model without mixing.
In our model, less than half of carbon is initially in CO, since considerable fraction of carbon is in H$_2$CO and CH$_4$.
Such low CO abundance has been recently observed in the inner envelopes of low-mass protostars, where $T>25$ K \citep{yildiz12},
and thus is indeed possible as an initial condition of the disk.
The timescale $\tau_{\rm CO}$ is inversely proportional to the ionization rate of He.
In our models, cosmic-ray ionization rate of H$_2$ is $5\times10^{-17}$ s$^{-1}$ referring to \citet{dalgarno06}.
If we set the cosmic-ray ionization rate of H$_2$ to be $1.3\times10^{-17}$ s$^{-1}$, which is commonly used value in astrochemical models, 
$\tau_{\rm CO}$ in the midplane is longer than 10$^6$ yr.
Note that the cosmic-ray ionization rate in disks remains highly uncertain in the current stage \citep{cleeves13}.
In the disk surface, on the other hand, X-ray ionization dominates over cosmic-ray ionization, and thus $\tau_{\rm CO}$ is shorter. 
A similar analysis is independently performed by \citet{bergin14}\footnote{The original manuscript of this article is submitted before 
the publication of Bergin et al. (2014).}.

In our simulations, the grain surface route is more important than the gas phase route in the regions with 
dust temperatures slightly higher than CO evaporation temperature.
In those regions, CO is mainly converted to CO$_2$ ice to be depleted from the gas phase.
In our model, $E_{\rm des}({\rm CO})$ is 1150 K and the energy barrier for diffusion is a half of the desorption energy.
Then at $T_{\rm d} = 25$ K, the resident time of CO on grain surface is a few years, which is long enough for CO to thermally hop over the whole grain surface
to find a reactant.

In Figure \ref{coms:fig:sink}, we can see the sharp drop of N$_2$ abundance after CO abundance considerably decreases ($\sim$10$^{-6}$).
Nitrogen depletion in the gas phase occurs by the following pathway: 
${\rm N_2}  \xrightarrow[]{{\rm H_3^+}} {\rm N_2H^+}  \xrightarrow[]{{\rm grain,\,e^{-}}} {\rm NH}$.
Once NH is formed, it is adsorbed onto dust grains, followed by subsequent hydrogenation to form NH$_3$ \citep{willacy07}.
Note that the above path is not efficient when CO is abundant;
most N$_2$H$^+$ reacts with CO and reforms N$_2$, due to higher proton affinity of CO than N$_2$.
Although N$_2$ is also destroyed by He$^+$, the products, N atom and N$^+$, mainly cycle back to N$_2$.

\subsubsection{Effect of Mixing}
Effect of turbulent mixing on carbon depletion in the gas phase varies with disk radius.
In the inner disks, where $\tau_{\rm CO} > \tau^{\rm des}_{\rm ice}$,
transport of ices from the midplane to the disk surface and photoreactions suppress the carbon depletion in the gas phase;
in the model with $\alpha_z=10^{-2}$, carbon depletion is not significant at $r\lesssim50$ AU in 10$^6$ yr (Figure \ref{coms:fig:sink_2d}),
although several \% of carbon still exists as carbon chain molecules at $r \gtrsim 20$ AU (see Figure \ref{coms:fig:r_vs_sigma_simple}).
In the midplane at $r\gtrsim50$ AU, on the other hand, vertical mixing increases atomic H abundance and enhances hydrogenation of CO on grain surfaces;
in the model with $\alpha_z=10^{-2}$, conversion of CO into CH$_3$OH occurs in the timescale of only several 10$^4$ yr at the midplane of $r\sim50$--100 AU, 
which is much shorter than $\tau_{\rm CO}$.
Transport of He$^+$ from the disk surface to the midplane is not important, since the destruction timescale of He$^+$ by CO is 
very short ($\tau_{\rm He^+} \approx (\beta n_{\rm CO})^{-1} \approx 10/n_{\rm CO}$ yr) compared to the dynamical timescale.
In the upper layers where stellar UV is heavily attenuated but Reaction (\ref{coms:react:co+he+}) is enhanced by X-ray ionization,
the sink mechanism is effective. But CO distribution is smoothed out by the mixing,
since the transport timescale over such layer is shorter than the depletion timescale.

\section{DISCUSSION \label{coms:sec:discussion}}
\subsection{Cometary molecules \label{coms:sec:comet}}
Table \ref{coms:table:comparison} summarizes observed abundances of selected species with respect to water 
in cometary coma \citep[][and reference therein]{mumma11} and low-mass protostellar envelopes \citep{oberg11}.
Table \ref{coms:table:comparison} also lists the column density ratio of icy molecules to water ice in the models with $\alpha_z=0$ and $10^{-2}$, 
and our initial abundances for disk chemistry (core).
We restrict our data to $r=20$--30 AU, where our model with $\alpha_z=10^{-2}$ reproduce the cometary HDO/H$_2$O ratio of 10$^{-3}$--10$^{-4}$, 
regardless of the initial HDO/H$_2$O ratio (Paper I).
We do not discuss CO and CH$_4$ here, because they are mostly in the gas phase in our model at $r=20$--30 AU, where $T_{\rm d}>25$ K.
In reality, H$_2$O ice matrix can trap these highly volatile species and prevent them from sublimating entirely
 at their sublimation temperatures \citep[e.g,][]{collings04}.
Our current model does not consider such trapping by water ice; 
the effect should be explored in future work.

As mentioned in the introduction, observed abundances of CH$_3$OH, NH$_3$ and CO$_2$ in comets are lower than the median abundances in the low-mass protostellar envelopes.
The model with $\alpha_z=10^{-2}$ reasonably reproduces abundances of NH$_3$ and CH$_3$OH in comets, while the model without mixing fails.
Although our initial abundances of NH$_3$ and CH$_3$OH are larger than those observed in ptorostellar envelopes by a factor of around two,
this discrepancy would not affect the result.
Then, our model suggests the possibility that the lower NH$_3$ and CH$_3$OH abundance in comets could be established in the solar nebula via the destruction 
and reformation of the interstellar ices.
Note, however, that we can not rule out the possibility that the ice composition in the parental core of the Solar nebula was methanol-poor and ammonia-poor 
compared to the median abundances in the low-mass protostellar envelopes.
The model with mixing also better reproduces the cometary C$_2$H$_6$ abundance.
On the other hand, our disk model both with and without mixing reasonably reproduce cometary CO$_2$ abundance.
One caution is that our initial composition is significantly CO$_2$-poor compared to that observed in ptorostellar envelopes.
The CO$_2$ abundance of the disk could be higher, if we start with higher CO$_2$ abundances.


Let us move on to complex organic molecules.
Complex molecules, such as HCOOCH$_3$, have been detected only in comet Hale-Bopp, which is the brightest comet in the past several decades.
The model without mixing reasonably reproduces the abundances of HCOOCH$_3$, CH$_3$CHO, and NH$_2$CHO, while the model abundance of HCOOH (and possibly CH$_3$OCH$_3$)
is lower than observations.
On the other hand, the model with mixing has significantly lower abundances than those in comets except for HCOOH; 
as mentioned in Section \ref{coms:sec:complex}, dust temperatures at disk surfaces of $r\lesssim60$ AU are too high to form complex molecules efficiently,
while in the midplane, higher abundance of atomic hydrogen than in the model without mixing suppress the formation of complex molecules.
Then, it is difficult to reproduce cometary abundances of both simple and complex molecules in a single model.

We note here that our model might overestimate the hydrogenation rate of radicals, and thus underestimate the abundance of complex molecules.
\citet{fuchs09} showed that hydrogen atoms can penetrate into top four monolayers of a CO ice matrix at maximum at the ice temperature of 12 K,
although the penetration depth slightly increase with temperature.
On the other hand, \citet{oberg09c} estimated that surface reactions are at most twice as efficient as bulk reactions in producing complex molecules.
These laboratory experiments suggests that radical-radical reaction could occur even in bulk ice layers, 
but hydrogenation by accreting H atom is effective only in the surface several layers.
In this study, we adopt a two-phase model, in which a layered structure of ice mantles is not considered, and the bulk of ices is chemically active, 
i.e., atomic hydrogen accreting onto grain surfaces can react with all species in the ice mantle, which hampers the formation of complex organic molecules.
Consideration of layered structure of ice and discrimination of the surface chemistry from bulk ice chemistry \citep{garrod13} are 
important to simulate synthesis of complex molecules in disks more accurately, which we will pursue in future work.

\subsection{Comparisons with previous models \label{coms:sec:comparison_model}}
SW11 studied effect of radial and vertical (2D) mixing on disk chemistry.
Here we briefly compare our results with SW11.
The comparison is not straightforward, however, because our model consider only the vertical mixing, 
and because both chemical and physical models are different between the two models.
Although SW11 utilized a reaction network based on \citet{garrod06} with various modifications like this work, 
adopted parameters, such as the diffusion-to-desorption energy ratio on grain surfaces, are different; 
we assume the ratio of 0.5, while it is 0.77 in SW11.
The disk physical model in SW11 is colder than that we use in this work; a maximum dust temperature in the midplane is less than 40 K in SW11.
Thus they do not report depletion of ice (e.g., H$_2$O and CH$_3$OH) found in the inner disks of our models.
In the following, we compare the effect of mixing on radial column density profiles 
in the regions where midplane dust temperature is less than 40 K ($r\gtrsim40$ AU in our model) between the two models.

SW11 found that the column densities of gaseous CH$_3$OH is enhanced by up to two orders of magnitude, 
while that of gaseous H$_2$CO is not strongly affected by mixing.
Our model predicts that the abundances of both species are enhanced through upward transport of ices containing CH$_3$OH, followed by photochemisty.
Although we do not have reasonable explanation for this discrepancy, we note that regardless of mixing strength the gas-phase H$_2$CO column density in SW11 is 
comparable to that in our model with $\alpha_z=10^{-2}$.
Abundant H$_2$CO in SW11 may have hidden the enhancement by mixing.
SW11 also found complex organic species, HCOOH ice and CH$_3$CHO ice, are more abundant in the model with mixing than in the model without mixing.
Our models confirm their results.


\subsection{Photodissociation rates of icy molecules \label{coms:dis:phd}}
Interaction between UV photons and ices is essential for chemistry in star-forming regions.
Details of the ice photolysisy are yet poorly understood.
In this subsection, we briefly discuss the effect of our assumptions in the calculations of photodissociation rates of ices on the results presented in Section \ref{coms:sec:result}.

In our fiducial models, we assumed that only the uppermost monolayer of the ice mantles can be dissociated as an outcome of photoabsorption, 
while photofragments immediately recombine in the deeper layers.
This assumption may lead to an underestimate of the photodissociation rates of ices, 
if the recombination probability of photofragments in the deeper layers is much less than unity.
As the opposite extreme, we have run the calculations with $N_{\rm p}=N_{\rm layer}$ instead of Equation (\ref{coms:eq:np}),
assuming that photofragments do not recombine even in the deepest layers.
If all oxygen is in water ice, $N_{\rm layer}$ is $\sim$50 monolayers in our models.
We found that the column densities of the molecules presented in Section \ref{coms:sec:result} are not sensitive to the assumption in $N_{\rm p}$; 
the difference of the column densities is mostly within a factor of three, regardless of the mixing strength.
CH$_3$OCH$_3$ ice in the model without mixing is most sensitive to $N_{\rm p}$;
the column density increases by about one order of magnitude relative to the fiducial model, due to the increased formation rate of CH$_3$ radical.
In the model with $\alpha_z=10^{-2}$, on the other hand, the column densities of HCOOCH$_3$ and CH$_3$OCH$_3$ ices slightly decrease relative to the fiducial model,
because the abundances of the simpler ices decrease in the disk surface due to the efficient photodissociation.

We also assumed that the photodissociation cross sections of icy molecules are the same as those of corresponding gaseous molecules,
except for water and CO$_2$ ices, in our fiducial models.
We have preformed a calculation in which the adopted cross sections of icy molecules are lowered by a factor of ten;
it also corresponds to the case that most of the photofragments recombine with the original partner even in the uppermost monolayer.
We confirmed that this does not affect the results of the simple molecules significantly.
On the other hand, in the model with $\alpha_z=10^{-2}$, column densities of HCOOCH$_3$ and CH$_3$OCH$_3$ ices decrease by up to 
a factor of five relative to the fiducial model,
due to the decreased radical formation rates in the disk surface.

\subsection{Effect of Grain Growth on the synthesis of complex organic molecules \label{coms:dis:grain}}
Observational studies of PPDs indicate that dust grains often have growth upto millimeter size or larger \citep[e.g.,][]{andrews05,ricchi10}.
In our fiducial models, a dust size distribution was assumed to be the interstellar one.
Grain growth reduces the total grain surface area, and thus reduces the rates of freeze-out and grain surface chemistry \citep[e.g.,][]{vasyunin11}.
In this subsection, we present models with larger grains than our fiducial models, and briefly discuss the effect of 
grain growth on the synthesis of complex organic molecules in turbulent disks.
More detailed discussions will be presented elsewhere.

The disk structure model we use here is similar to that of \citet{aikawa06}.
In the model, the gas temperature, dust temperature and density distributions of the disk are calculated self-consistently, 
considering various heating and cooling mechanisms.
A dust size distribution is the power-law, $df(a)/da \propto a^{-3.5}$, with the 
minimum and maximum dust radii of 0.01 $\mu$m and 1 mm, respectively.
Dust sedimentation is not considered, and dust-to-gas mass ratio is set to be 0.01 in the whole disk.
Then, the total surface area of dust is lower by a factor of ten than in our fiducial disk model (Section \ref{coms:sec:phys_disk}).
In the chemical simulation, we reduce the total surface area of grains accordingly.

Figure \ref{coms:fig:1mm} shows the physical parameters and abundances of HCOOCH$_3$ ice and CH$_3$OCH$_3$ ice 
in the models with $\alpha_z = 0$ and $\alpha_z = 10^{-2}$ at $t = 10^6$ yr.
The physical structure of the disk is significantly different from our fiducial model; 
the disk temperature is lower, and UV penetrates deeper into the disk.
Yet we confirmed that COMs are mainly formed at $A_V =$ a few mag in the model with $\alpha_z=10^{-2}$, as in the case of our fiducial model.
While the regions with moderate COMs abundance ($x_i \gtrsim 10^{-8}$) are small in the model without mixing, 
the vertical mixing expands such regions both in the vertical direction and to outer radii.
The impact of vertical mixing is greater in the large grain model than in our fiducial model; 
the UV photons penetrates to deeper layers of the disk, where the higher density of gas and dust make the COMs formation efficient.
Compared with our fiducial disk model, the disk model with large grains has lower dust temperatures in the midplane, 
which suppresses the formation of COMs at the outer radii ($r\gtrsim20$ AU) in the model without mixing.
 
\section{Conclusion \label{coms:sec:conclusion}}
We have investigated ice chemistry in turbulent protoplanetary disks surrounding a T Tauri star, focusing on carbon and nitrogen bearing molecules.
We have solved chemical rate equations with the diffusion term, mimicking the turbulent mixing in the vertical direction.
Turbulence brings ice-coated dust grains from the midplane to the disk surface, where 
the strong UV radiation field and higher dust temperature activate ice chemistry.
Transport of atomic hydrogen from the disk surface to the midplane also plays an important role for ice chemistry.
Our main conclusions are as follows.

\begin{enumerate}
{\item Upward transport of ices decreases the abundance of saturated ice molecules, CH$_3$OH and NH$_3$, in the inner disks ($r\lesssim30$ AU), 
because warm dust temperatures prohibit their reformation via the hydrogenation on grain surfaces.
At $r\sim$20--30 AU the resultant ice composition is methanol-poor and ammonia-poor compared to the initial composition, 
because water ice abundance decreases only at $r\lesssim 20$ AU.
This contrasts with the model without mixing, where the ice composition is methanol-rich and ammonia-rich compared to the initial composition.}

{\item In the model with $\alpha_z \lesssim 10^{-3}$, complex organic molecules are mainly formed in the midplane.
We found that turbulent mixing affects synthesis of complex organic molecules in the two ways:
(1) transport of ices from the  midplane to the disk surface and
(2) transport of atomic hydrogen from the disk surface to the midplane.
The former enhances the formation of complex molecules in the disk surface, 
while the latter enhances the hydrogenation of radicals and thus suppresses the formation of complex molecules in the midplane.
As a result, complex molecules are mainly formed in the disk surface when mixing is strong ($\alpha_z=10^{-2}$).
It means that complex molecules can be formed even in the outer disks, where 
dust temperatures in the midplane are too low for their formation.
Vertical mixing thus expands the distribution of complex molecules both in the vertical direction and to outer radii compared with that in the non-turbulent case.
This is true both in the model with interstellar dust grains and in the model with larger dust grains (Section 4.4).
}

{\item It is known that comets are depleted in simple saturated species relative to water compared to the median abundances
of the ices in low-mass protostellar envelopes.
The model with $\alpha_z=10^{-2}$ reasonably reproduces the abundances of simple molecules in comets,
while the model without mixing fails.
Our results suggest that the observed deficiency of simple saturated species 
in comets might imply the destruction and reformation of the ices in the solar nebula.
However, the model without mixing better reproduces the abundances of complex molecules in comets.
Then, it is difficult to reproduce cometary abundances of both simple and complex molecules in a single model.
To resolve this issue, we need to introduce layered structure of ice mantles in our model,
considering that atomic hydrogen cannot penetrate deeply in ice mantles, according to laboratory experiments.
}

{\item Upward transport of ices enriches icy molecules in the disk surface, where destruction by photoreactions and supply from the deeper layers are balanced.
It also affects the abundances of gaseous species through photodesorption.
In the model with $\alpha_z=10^{-2}$, radial profiles of column densities of H$_2$CO and CH$_3$OH in the gas phase are ring-like. 
The inner edge of the ring is determined by the efficiency of CO hydrogenation on grain surfaces.
The analytical expressions of radial profiles of CH$_3$OH column density are given in Appendix B, which would be useful for 
analyzing (near) future radio observations by ALMA.
}

{\item In the disk chemical models, it often occurs that carbon and nitrogen are locked in ices even in the regions 
where dust temperature is greater than the sublimation temperatures of CO and N$_2$ ($\gtrsim$20--25 K).
Recent observation by \citet{favre13} suggests that such a sink mechanism is really working in TW hya.
We found that characteristic timescale of carbon depletion in the gas phase ($\tau_{\rm CO}$) is given by Equation (\ref{coms:eq:t_co}),
and once CO abundance drops after $\tau_{\rm CO}$, depletion is accelerated.
Nitrogen depletion occurs only after CO abundance considerably decreases ($\sim$10$^{-6}$).
Effect of turbulent mixing on the chemistry varies with disk radius and height.
}

\end{enumerate}



\acknowledgments
We thank Hideko Nomura for providing disk physical models.
We also thank the anonymous referee for his/her useful comments, which improve our manuscript considerably.
This work was partly supported by a grant-in-aid for Scientific Research (23103004, 23540266) of the Ministry of Education, 
Culture, Sports, Science and Technology of Japan (MEXT).
K.F. is supported by the Research Fellowship from the Japan Society for the Promotion of Science (JSPS).

\appendix
\section{Evaporation temperature}
Evaporation temperature of icy species can be defined as the dust temperature at which adsorption timescale of gaseous species onto dust grains ($\tau_{\rm ads}$) and 
thermal desorption timescale of the corresponding icy species ($\tau_{\rm des}$) balance:
\begin{align}
\tau_{\rm ads} &= \tau_{\rm des},\\
\tau_{\rm ads} &= (\pi a^2 n_d v_{\rm th})^{-1},\\
\tau_{\rm des} &= \nu^{-1} \exp(E_{\rm des}/T_{\rm d}),
\end{align}
where $v_{\rm th}$ and $E_{\rm des}$ are the thermal velocity of the gaseous species and desorption energy, respectively. 
The pre-exponential factor $\nu$ is the vibrational frequency of the molecule in the binding site, and is evaluated by using the
harmonic oscillator strength \citep[see Equation (3) of][]{hasegawa92}.
The typical value of $\nu$ is 10$^{12}$--10$^{13}$ s$^{-1}$ \citep{hasegawa92}.
Evaporation temperatures listed in Table \ref{coms:table:tevap} are calculated at the gas density of $10^6$ cm$^{-3}$ and 10$^{12}$ cm$^{-3}$, 
assuming dust and gas temperatures are same.
Note that the evaporation temperature weakly depends on gas density and molecular mass.

\section{Analytical expressions of column density profiles of gaseous methanol}
In this section, we give analytical expressions of the column density profile of gaseous CH$_3$OH in protoplanetary disks,
which reasonably reproduce our numerical results both in the model with $\alpha_z=0$ and 10$^{-2}$ as shown in Figure \ref{coms:fig:r_vs_sigma_gas}.
We focus on CH$_3$OH, since no efficient formation pathway is known for CH$_3$OH in the gas phase.
However, the following description can be easily extended and applied to other molecules.

\begin{figure}
\epsscale{1.0}
\plotone{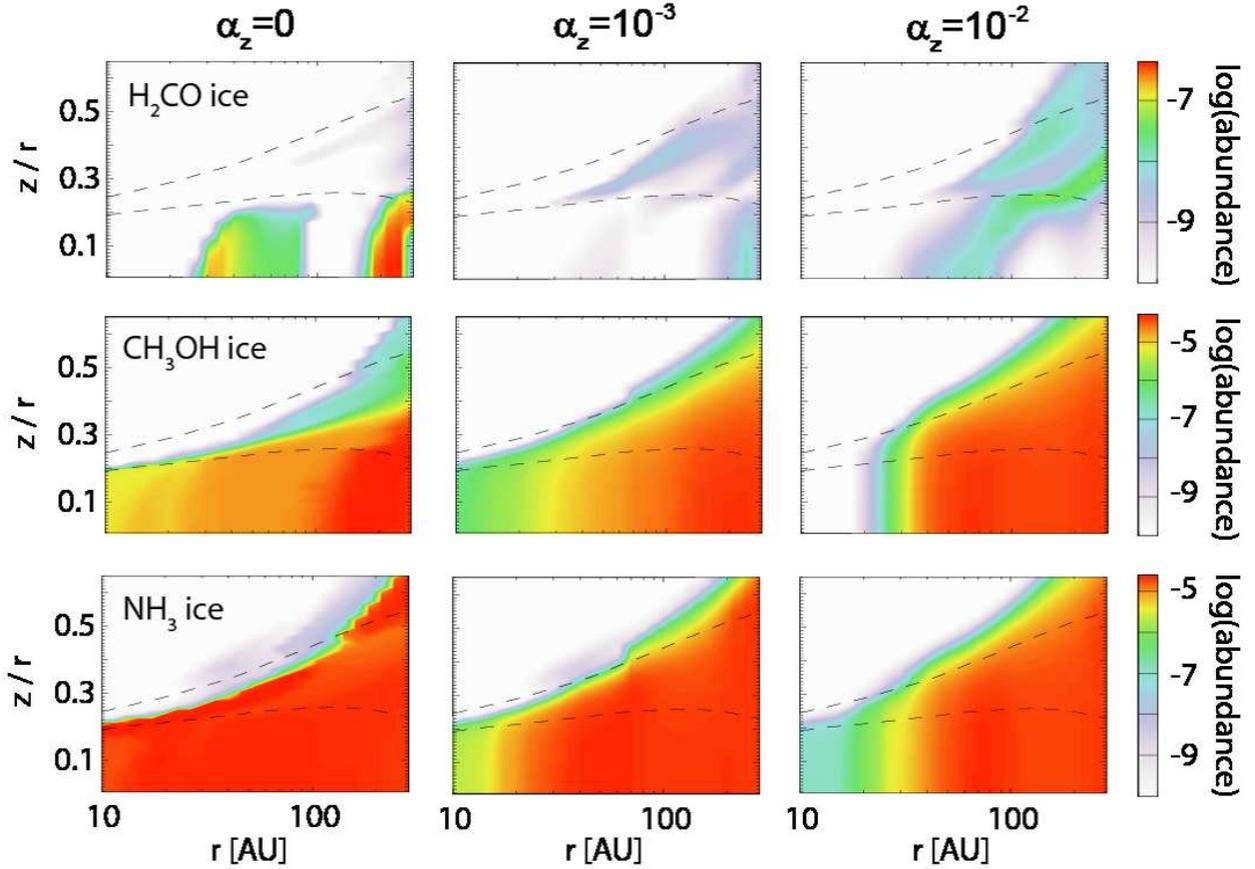}
\caption{Spatial distributions of fractional abundances of selected molecules with respect to hydrogen nuclei.
The left, middle and right columns are models with $\alpha_z = 0$, 10$^{-3}$, and 10$^{-2}$, respectively.
The dashed lines indicate the vertical visual extinction of unity and 10.
\label{coms:fig:ab_simple}}
\end{figure}

\begin{figure}
\epsscale{1.0}
\plotone{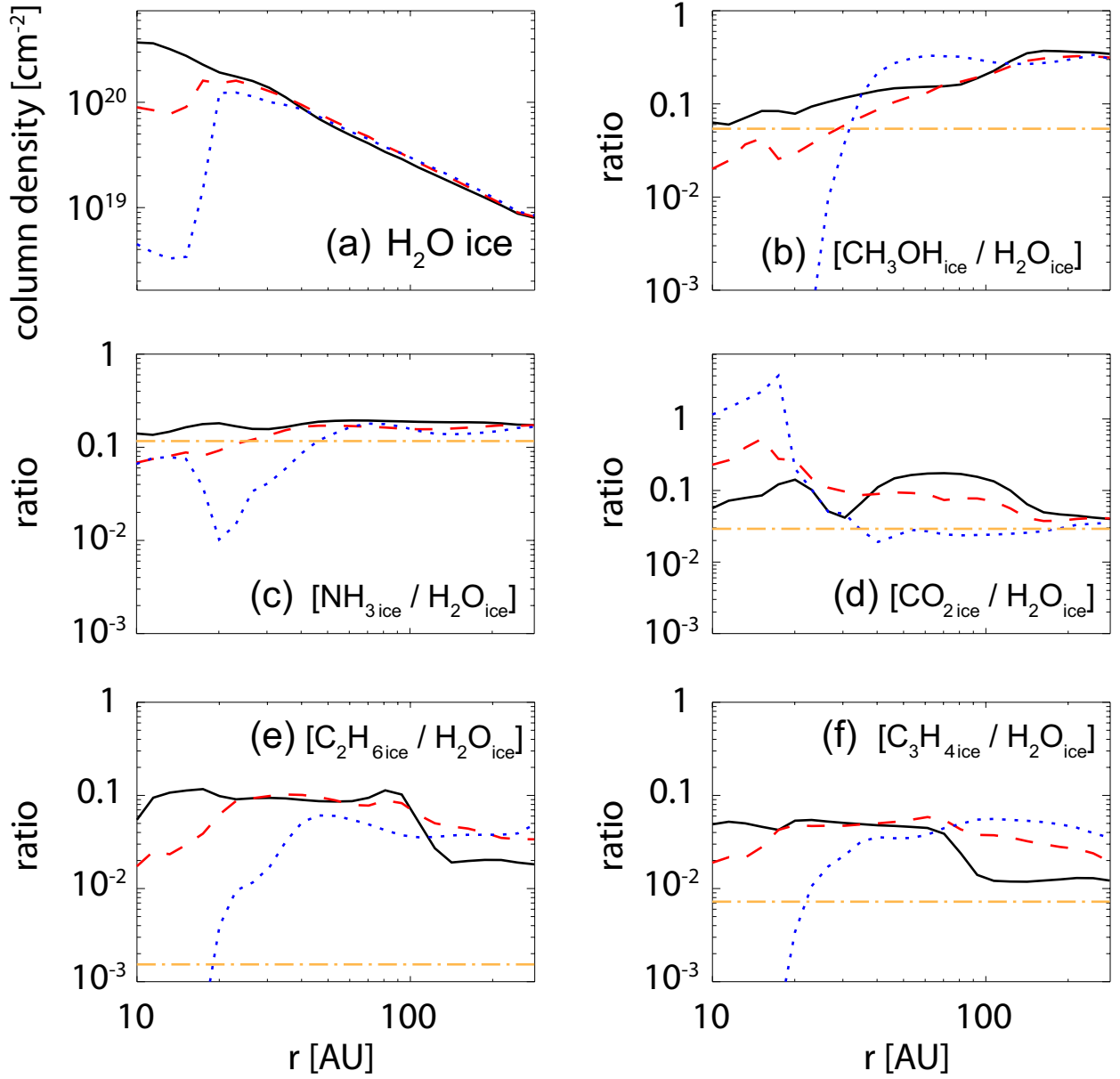}
\caption{Radial profiles of water ice column density (a).
Panels (b)-(f) show column densities of selected icy molecules normalized by water ice column density. 
The solid, dashed, and dotted lines represent the model with $\alpha_z = 0$, $10^{-3}$, and $10^{-2}$, respectively. The horizontal dash-dotted lines show
the initial molecular abundances with respective to water (see Table \ref{coms:table:initial}).
\label{coms:fig:r_vs_sigma_simple}}
\end{figure}

\begin{figure}
\epsscale{1.0}
\plotone{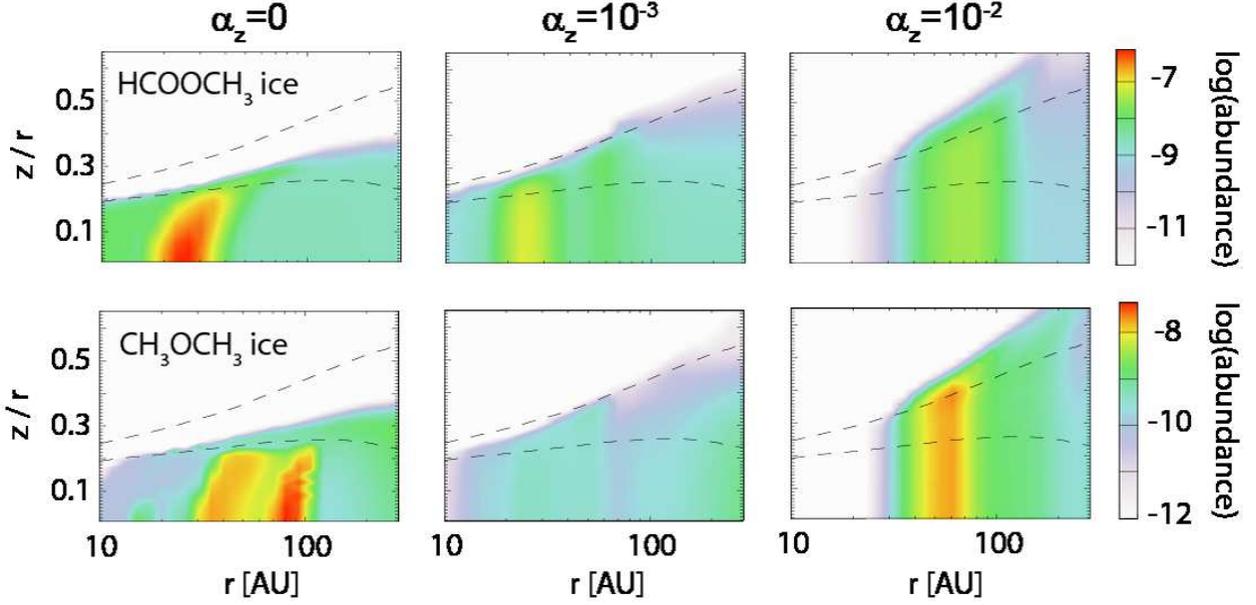}
\caption{Spatial distributions of HCOOCH$_3$ ice (top) and CH$_3$OCH$_3$ ice (bottom) abundances.
Other details are the same as Figure \ref{coms:fig:ab_simple}.
\label{coms:fig:ab_com}}
\end{figure}

\begin{figure}
\epsscale{0.7}
\plotone{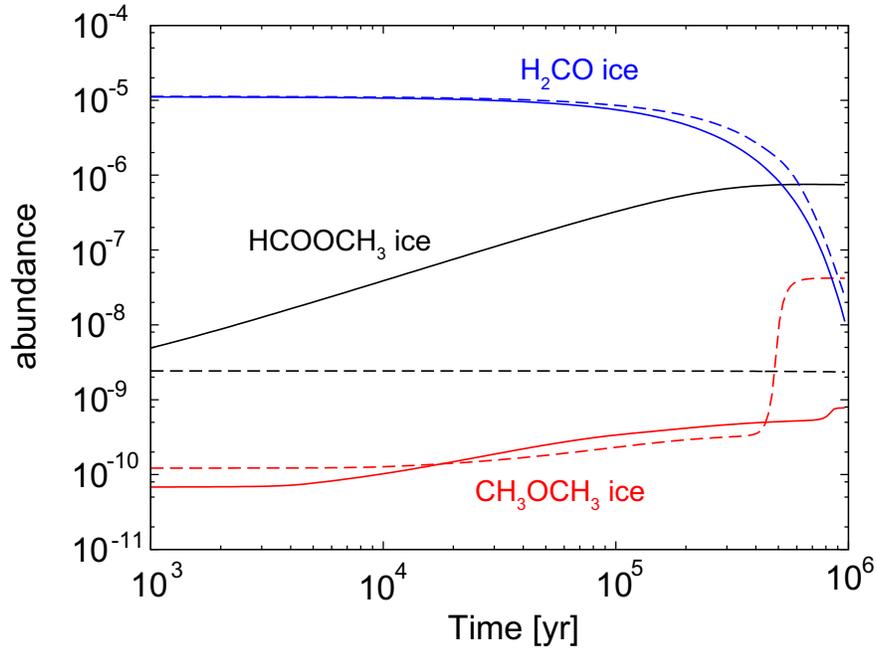}
\caption{Temporal variations of abundances of HCOOCH$_3$ ice (black), CH$_3$OCH$_3$ ice (red), and H$_2$CO ice (blue) in the model without mixing 
at the midplane of $r=25$ AU (solid) and $r=80$ AU (dashed). \label{coms:fig:t_vs_ab_nd}}
\end{figure}

\begin{figure}
\epsscale{1.0}
\plotone{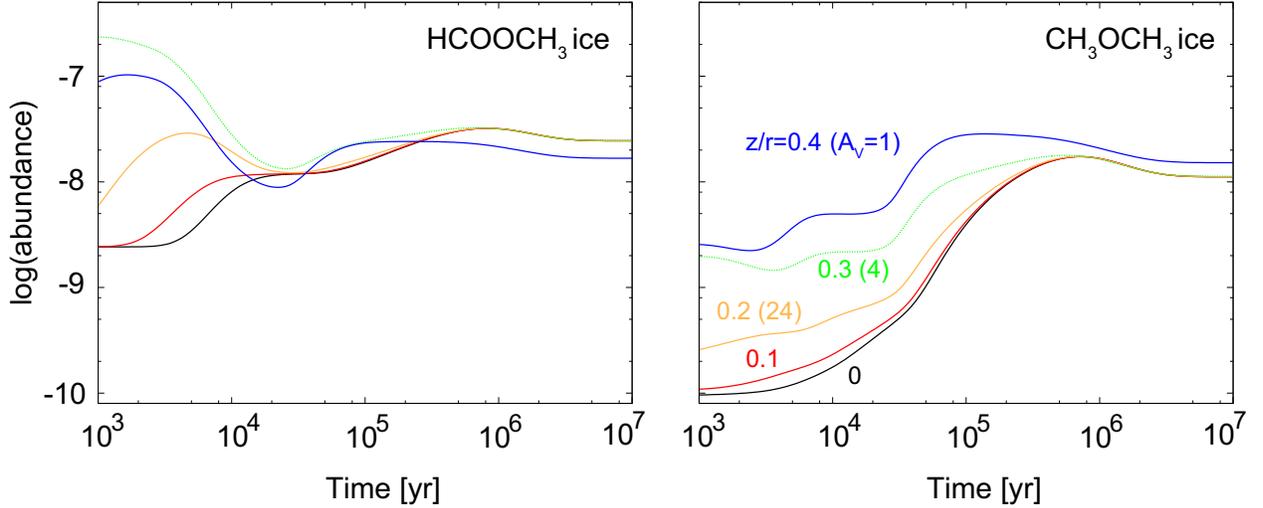}
\caption{Temporal variations of abundances of HCOOCH$_3$ ice (left) and CH$_3$OCH$_3$ ice (right) in the model with $\alpha_z = 10^{-2}$ and $r=60$ AU
at various heights from the midplane. The values in bracket are the vertical visual extinction from the disk surface. \label{coms:fig:t_vs_ab_sd}}
\end{figure}

\begin{figure}
\epsscale{1.0}
\plotone{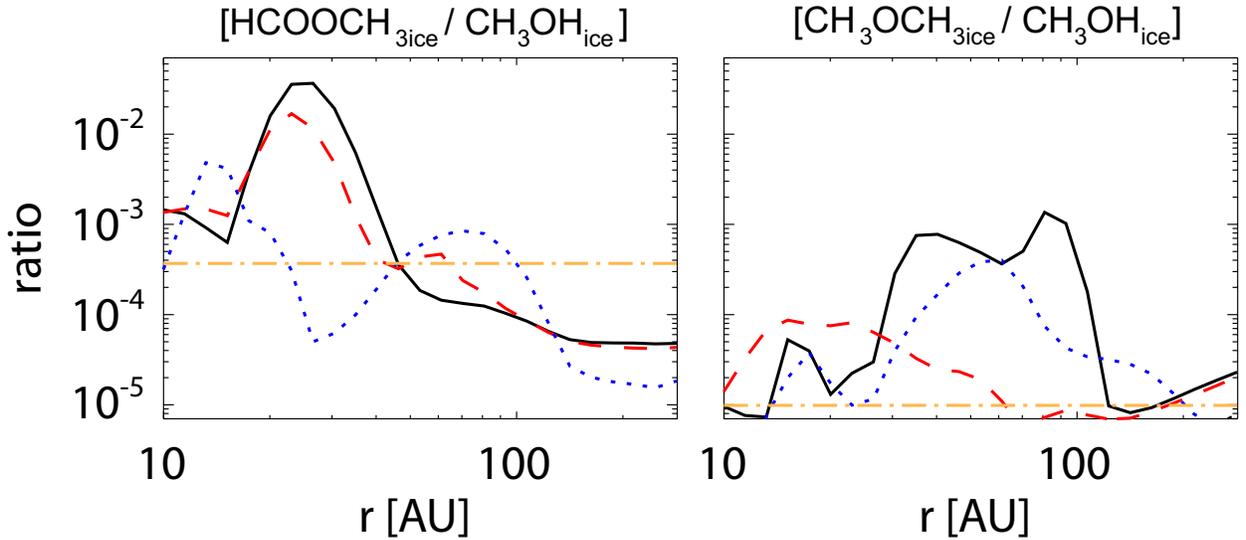}
\caption{Radial profiles of column densities of HCOOCH$_3$ ice (left) and CH$_3$OCH$_3$ ice (right), normalized by CH$_3$OH ice column density. 
The solid, dashed, and dotted lines represent the model with $\alpha_z = 0$, $10^{-3}$, and $10^{-2}$, respectively.
The horizontal dash-dotted lines show the initial HCOOCH$_3$/CH$_3$OH and CH$_3$OCH$_3$/CH$_3$OH abundance ratios 
(see Table \ref{coms:table:initial}). \label{coms:fig:r_vs_sigma_com}}
\end{figure}

\begin{figure}
\epsscale{1.0}
\plotone{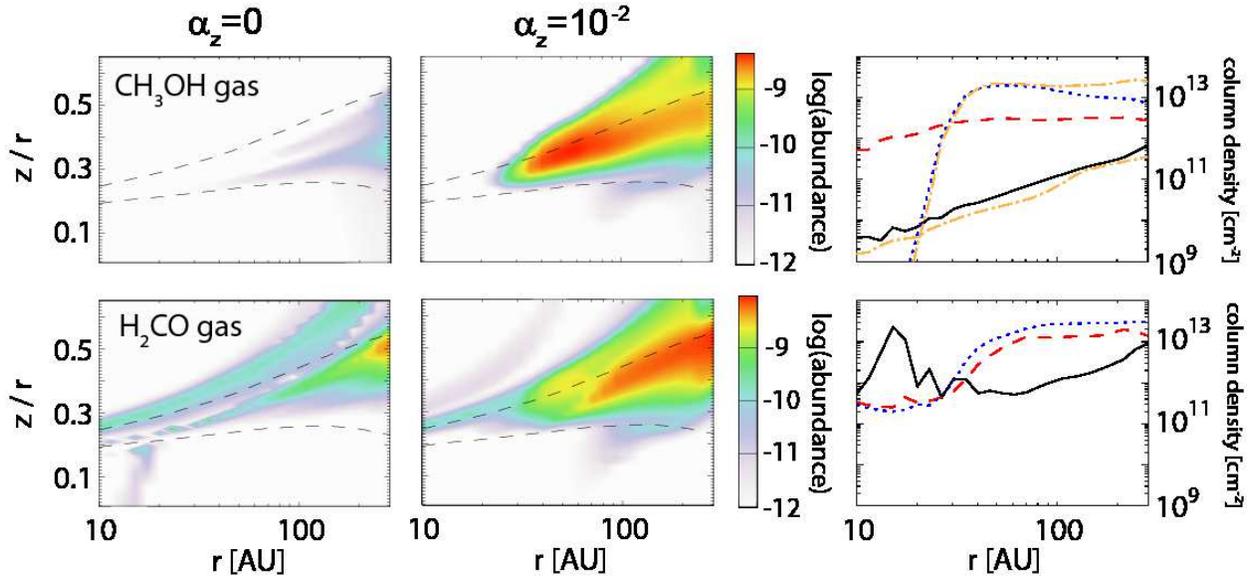}
\caption{Spatial distributions of the abundances (left and middle) and column densities (right) of gaseous CH$_3$OH and H$_2$CO.
The solid, dashed, and dotted lines in right panels represent the model with $\alpha_z = 0$, $10^{-3}$, and $10^{-2}$, respectively,
while the dash-dotted lines depict the column density profiles given by Equations (\ref{coms:eq:N_gas_nd}) and (\ref{coms:eq:N_gas_sd}) in Appendix B, 
using $n_i^{\rm s}/n_{\rm ice}$ in the numerical models.
\label{coms:fig:r_vs_sigma_gas}}
\end{figure}

\begin{figure}
\epsscale{1.0}
\plotone{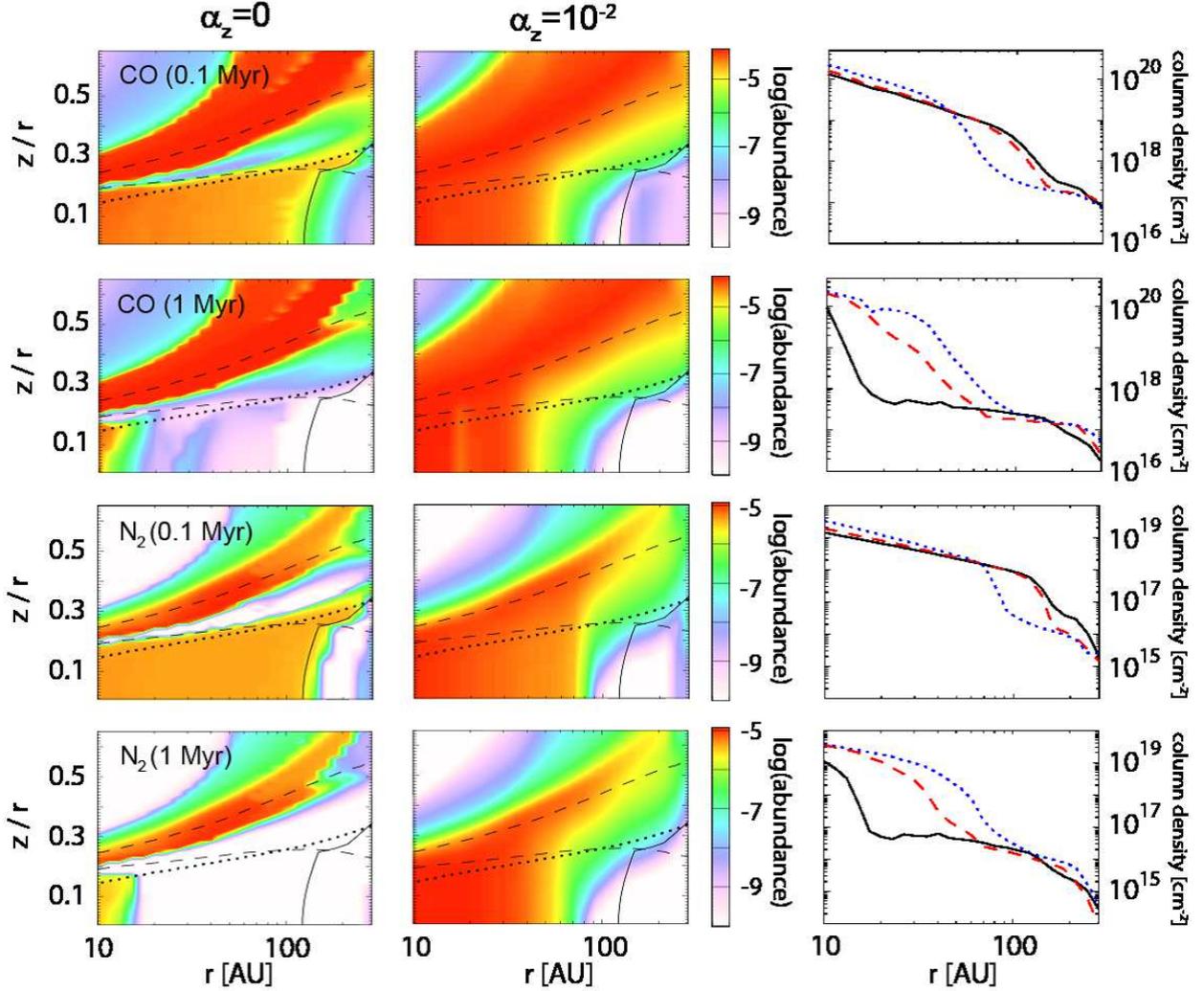}
\caption{Spatial distributions of the abundances (left and middle) and column densities (right) of CO and N$_2$ at 10$^5$ yr and 10$^6$ yr.
In left and middle panels, the solid lines indicate dust temperature of 24 K, which nearly corresponds to sublimation temperature of CO.
while the dashed lines indicate the vertical visual extinction of unity and 10. The dotted lines indicate the boundary above which X-ray is the dominant
ionization source, while the cosmic-ray is the dominant ionization source in the lower layers.
\label{coms:fig:sink_2d}}
\end{figure}

\begin{figure}
\epsscale{0.7}
\plotone{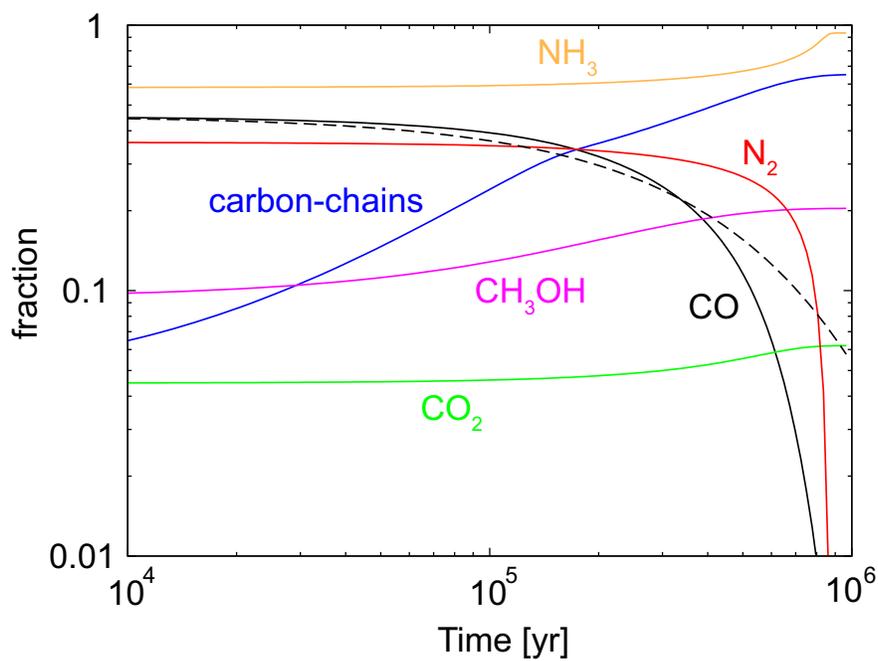}
\caption{Temporal variations of a fraction of elemental carbon and nitrogen in the form of selected species at the midplane of $r=25$ AU in the model without mixing.
Dust temperature is $\sim$40 K.
The dashed line depicts the temporal variation given by $f_{\rm CO}\exp(-t/\tau_{\rm CO})$, where $f_{\rm CO}$ is the initial fraction of carbon in CO. 
\label{coms:fig:sink}}
\end{figure}

\begin{figure}
\epsscale{1.0}
\plotone{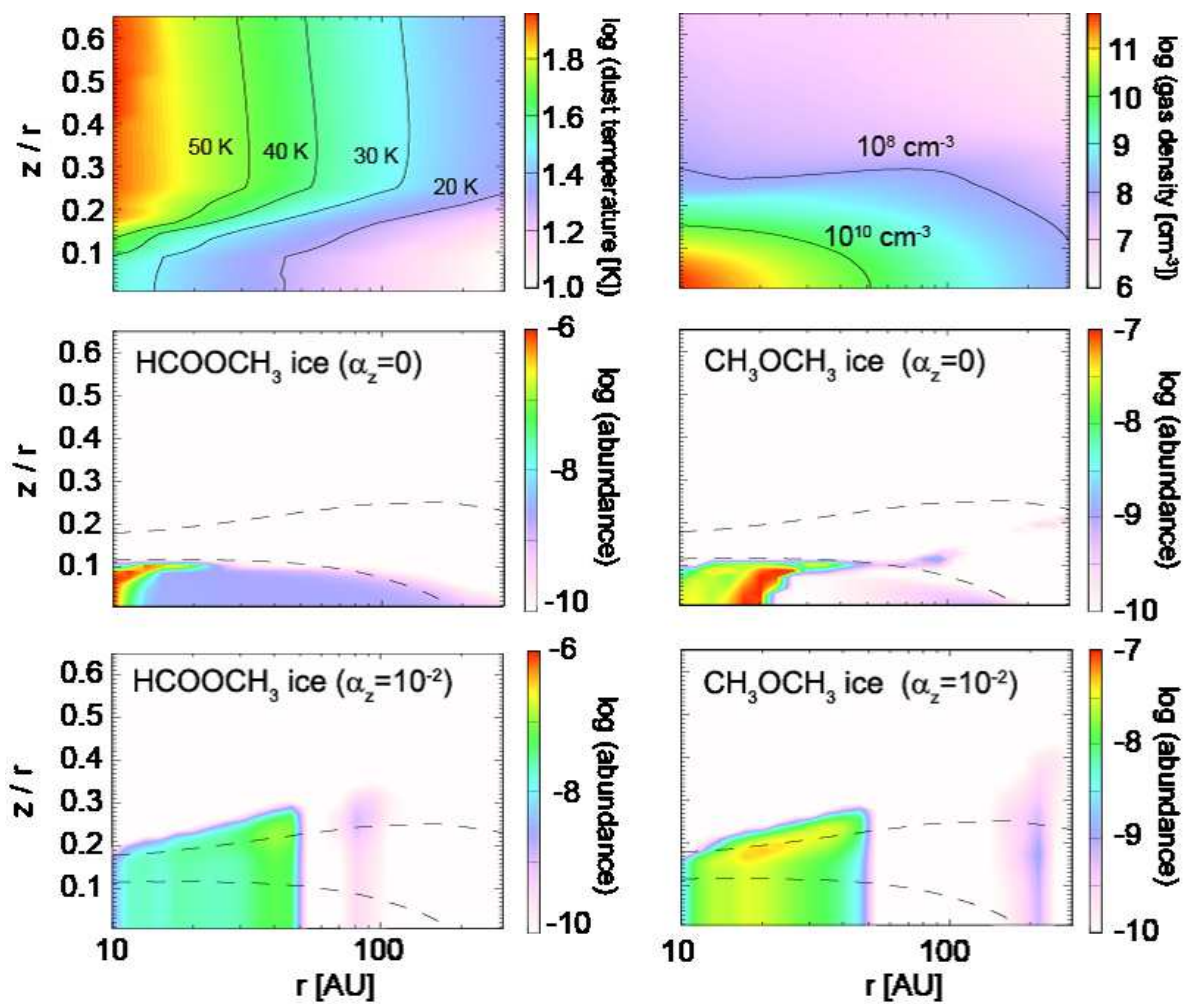}
\caption{Spatial distributions of physical parameters (top panels) and abundances of HCOOCH$_3$ ice (middle panels) 
and CH$_3$OCH$_3$ ice (bottom panels) at $t = 10^6$ yr in the model with large dust grains. Dashed lines indicate the vertical visual extinction of unity and 10.
\label{coms:fig:1mm}}
\end{figure}

In the model without mixing, CH$_3$OH ice predominantly exists near the midplane (Figure \ref{coms:fig:ab_simple}).
In that case, the number density of gaseous CH$_3$OH near the midplane is evaluated from the balance between the photodesorption by cosmic-ray induced photons and adsorption onto dust grains:
\begin{equation}
n^{\rm g}_{\rm CH_3OH} = 10^{-4}\left(\frac{n_{\rm CH_3OH}^{\rm s}/n_{\rm ice}}{0.1}\right)\left(\frac{F_{\rm CRUV}}{10^4\,{\rm cm^2\,s^{-1}}}\right)
\left(\frac{Y}{10^{-3}}\right)\left(\frac{v_{\rm th}}{10^4\,{\rm cm\,s^{-1}}}\right)^{-1}.
\end{equation}
The number density of gaseous CH$_3$OH is independent of the number density of gas.
Since CH$_3$OH ice is abundant only at $z/r \lesssim 0.3$ in the model without mixing,
the radial column density profile of gaseous CH$_3$OH is evaluated by
\begin{align}
N^{\rm g}_{i} & \approx n^{\rm g}_{\rm CH_3OH} \times 0.3r \\
              & \approx 10^{11}\left(\frac{n_{\rm CH_3OH}^{\rm s}/n_{\rm ice}}{0.1}\right) \left(\frac{F_{\rm CRUV}}{10^4\,{\rm cm^2\,s^{-1}}}\right)
\left(\frac{Y}{10^{-3}}\right)\left(\frac{v_{\rm th}}{10^4\,{\rm cm\,s^{-1}}}\right)^{-1}
\left(\frac{r}{300\,{\rm AU}}\right)\left(\frac{z/r}{0.3}\right) \,\,{\rm cm^{-2}}.\label{coms:eq:N_gas_nd}
\end{align}
Equation (\ref{coms:eq:N_gas_nd}) shows that the column density is proportional to radius.
The radial dependence can be steeper, if $n_{\rm CH_3OH}^{\rm s}/n_{\rm ice}$ increases with radius.
The dot-dashed line in Figure \ref{coms:fig:r_vs_sigma_gas} shows the analytical value with $n_{\rm CH_3OH}^s/n_{\rm ice}$ adopted from our numerical model.

On the other hand, in the model with mixing, CH$_3$OH ice exists even in the disk surface.
As an extreme case, let us assume that CH$_3$OH ice exits in the whole disk, and photodissociation is the main destruction path of gaseous CH$_3$OH in the disk surface.
As mentioned in Section \ref{coms:sec:comparison_obs}, the latter assumption is not valid at $r\gtrsim200$ AU in our model.
But errors introduced by the latter assumption is not so large as seen in Figure \ref{coms:fig:r_vs_sigma_gas}.
Considering Equation (\ref{coms:eq:x_sd}), the radial column density profile of gaseous CH$_3$OH is evaluated by $N^{\rm g}_{\rm CH_3OH} \approx x^{\rm g}_{\rm CH_3OH} \times N_{\rm H}^{\rm crit}$, 
where $N_{\rm H}^{\rm crit}$ is the column density of hydrogen nuclei measured from the disk surface to the height ($\sim4\times10^{21}$ cm$^{-2}$ in our model), 
below which ion-neutral reactions or adsorption onto dust grains becomes dominant destruction path of gaseous CH$_3$OH instead of photodissociation:
\begin{equation}
N^{\rm g}_{\rm CH_3OH} \approx 2\times 10^{13}\left(\frac{n_{\rm CH_3OH}^{\rm s}/n_{\rm ice}}{0.1}\right)\left(\frac{\pi a^2 x_d N_{\rm H}^{\rm crit}}{2}\right)
\left(\frac{Y}{10^{-3}}\right)\left(\frac{\sigma_{\rm Ly\alpha}}{10^{-17}\,{\rm cm^2}}\right)^{-1} \,\,{\rm cm^{-2}}.\label{coms:eq:N_gas_sd}
\end{equation}
Equation (\ref{coms:eq:N_gas_sd}) shows that the column density is dependent on radius only through $n_{\rm CH_3OH}^{\rm s}/n_{\rm ice}$.
Then the radial profile in the model with mixing is flattened compared to that in the model without mixing.
Equations (\ref{coms:eq:N_gas_nd}) and (\ref{coms:eq:N_gas_sd}) also explain the order-of-magnitude enhancement of the column density via the vertical ice transport.

\clearpage



\clearpage
\begin{table}
\begin{center}
\caption{Evaporation Temperature of Selected Icy Species.\label{coms:table:tevap}}
\begin{tabular}{ccc}
\tableline\tableline
Species & $E_{\rm des}$ (K)\footnotemark[1] & $T_{\rm evap}$ (K)\footnotemark[2] \\
\tableline
H             & 600   & 11--15\\
N$_2$         & 1000  & 19--26\\
CO            & 1150  & 23--31\\
CH$_3$        & 1175  & 22--30\\
HCO           & 1600  & 30--41\\
H$_2$CO       & 2050  & 39--52\\
CO$_2$        & 2650  & 50--68\\
CH$_3$OCH$_3$ & 3150  & 59--80\\
C$_2$H$_6$    & 4387  & 83--110\\
CH$_3$O       & 5084  & 96--130\\
CH$_3$OH      & 5530  & 100--140\\
NH$_3$        & 5534  & 100--140\\
H$_2$O        & 5700  & 110--150\\
HCOOCH$_3$    & 6300  & 120--160\\

\tableline
\end{tabular}
\tablenotetext{1}{Desorption energy of icy species on water ice. The values are taken from \citet{garrod06} except for H atom; it is from \citet{al-halabi07}.}
\tablenotetext{2}{Evaporation temperature of icy species when the gas density ranges from 10$^6$ cm$^{-3}$ (left) to 10$^{12}$ cm$^{-3}$ (right).
See Appendix A for more information.}
\end{center}
\end{table}

\begin{table}
\begin{center}
\caption{Initial Abundances of Assorted Species With Respect To Hydrogen Nuclei for the Disk Chemistry.\label{coms:table:initial}}
\begin{tabular}{cccc}
\tableline\tableline
Species & Abundance\footnotemark[1] & Species & Abundance \\
\tableline
H$_2$O  & 1.2(-4)       &  CO     & 3.6(-5)\\
CO$_2$ & 3.5(-6)        & CH$_4$ & 1.5(-5)\\
H$_2$CO & 1.2(-5)       & CH$_3$OH& 6.5(-6)\\
HCOOCH$_3$& 2.4(-9)     & CH$_3$OCH$_3$& 6.4(-11)\\
N$_2$  & 4.5(-6)        & NH$_3$ & 1.4(-5)\\
\tableline
\end{tabular}
\tablenotetext{1}{$a(-b)$ means $a\times10^{-b}$.}
\end{center}
\end{table}

\begin{table}
\begin{center}
\caption{Molecular Abundances of Cometary Coma and Low-Mass Protostellar Envelopes \label{coms:table:comparison}}
\begin{tabular}{cccccc}
\tableline\tableline
Species      & Comet\footnotemark[1] & Protostar\footnotemark[2] & \multicolumn{3}{c}{Model\footnotemark[3]}   \\ 
\cline{4-6}
             &         &          & ($\alpha_z=0$)   & ($\alpha_z=10^{-2}$) & Core               \\
\tableline
H$_2$O       & 100     & 100      & 100    & 100     & 100          \\
CO$_2$       & 2--30   & 29       & 4--14 & 4--19    & 3           \\
NH$_3$       & 0.2--1.4&  5       & 16--18 & 1--4    & 12           \\
CH$_3$OH     & 0.2-7&  3       &  8--12 & 0.004--4&  6           \\
C$_2$H$_6$   & 0.1-1.7 &  ...     &  9--10 & 0.4--1.6&  0.2           \\
HCOOCH$_3$   & 0.08    &  ...     & 0.1--0.2 & (0.04--$4)\times10^{-4}$&  0.002       \\
CH$_3$OCH$_3$& $<0.5$  &  ...     & (0.1--$3)\times10^{-3}$ & (0.007--$4)\times10^{-4}$ & $5\times10^{-5}$       \\
HCOOH        & 0.09    &  ...     & (0.8--$3)\times10^{-3}$ & 0.1--0.2  & $8\times10^{-4}$       \\
CH$_3$CHO    & 0.02    &  ...     & 0.04 & $<10^{-5}$  & 0.05       \\
NH$_2$CHO    & 0.015   &  ...     &  0.07--0.1 & (0.2--$3)\times10^{-3}$  & 0.08      \\
\tableline
\end{tabular}
\tablenotetext{1}{The values are taken from \citet{mumma11}.}
\tablenotetext{2}{The median ice composition in low-mass protostellar envelopes \citep{oberg11}.}
\tablenotetext{3}{Molecular column density normalized by that of water at $r=20$--30 AU, at $t=10^6$ yr. "Core" represents the initial composition for disk chemistry.}
\end{center}
\end{table}

\end{document}